\documentclass[preprint]{aastex}
\input epsf
\newcommand{\reff}{\mbox{$R_{\rm e}$}}
\newcommand{\ishape}{{\tt ishape}}
\newcommand{\mksynth}{{\tt mksynth}}
\newcommand{\tinytim}{{\tt tinytim}}
\newcommand{\feh}{\mbox{[Fe/H]}}
\newcommand{\vi}{\mbox{$V\!-\!I$}}
\newcommand{\vio}{\mbox{$(V\!-\!I)_0$}}

\newcommand{\bko}{\mbox{$(B\!-\!K)_0$}}
\newcommand{\bi}{\mbox{$B\!-\!I$}}
\newcommand{\jk}{\mbox{$J\!-\!K$}}
\newcommand{\jko}{\mbox{$(J\!-\!K)_0$}}
\newcommand{\bio}{\mbox{$(B\!-\!I)_0$}}

\newcommand{\msun}{\mbox{M$_{\odot}$}}
\newcommand{\lsun}{\mbox{L$_{\odot}$}}
\newcommand{\mto}{\mbox{$m^{\rm TO}_V$}}
\newcommand{\dmto}{\mbox{$\Delta m^{\rm TO}_V$}}
\begin{document}
\title{Properties of Globular Cluster Systems in Nearby Early-type Galaxies
\footnote{Based on observations with the NASA/ESA Hubble Space
 Telescope, obtained at the Space Telescope Science Institute, which is
 operated by the Association of Universities for Research in Astronomy,
 Inc. under NASA contract No.  NAS5-26555.}
}
\author{S{\o}ren S. Larsen and Jean P. Brodie  
  \affil{UCO / Lick Observatory, UC Santa Cruz, USA} 
  \email{soeren@ucolick.org, brodie@ucolick.org}
\and
  John P. Huchra 
  \affil{Harvard-Smithsonian Center for Astrophysics, USA} 
\and
   Duncan A. Forbes 
   \affil{Astrophysics \& Supercomputing, Swinburne University, 
          Hawthorn VIC 3122, Australia} 
\and
  Carl Grillmair 
  \affil{SIRTF Science Center, California Institute of Technology, USA} 
}

\begin{abstract}
  We present a study of globular clusters (GCs) in 17 relatively nearby
early-type galaxies, based on deep F555W and F814W images from the 
Wide Field / Planetary
Camera 2 (WFPC2) on board the Hubble Space Telescope (HST). A detailed
analysis of color distributions, cluster sizes and luminosity functions 
is performed and compared with GCs in the Milky Way.  In nearly all cases,
a KMM test returns a high confidence level for the hypothesis that a sum 
of two Gaussians provides a better fit to the observed color distribution 
than a single Gaussian, although histograms of the \vio\ distribution are
not always obviously bimodal.  The blue and red peak colors returned by 
the KMM test are \emph{both} found to correlate with absolute host 
galaxy $B$ band magnitude and central velocity dispersion (at about the
$2-3\sigma$ level), but we see no clear correlation with host galaxy \vi\ 
or \jk\ color.  Red GCs are generally smaller than blue GCs
by about 20\%. The size difference is seen at all radii and within sub-bins 
in \vio\ color, and exists also in the Milky Way and Sombrero (M104) spiral
galaxies. Fitting $t_5$ functions to the luminosity functions of blue and 
red GC populations
separately, we find that the $V$-band turn-over of the blue GCs is brighter 
than that of the red ones by about 0.3 mag on the average, as expected if
the two GC populations have similar ages and mass distributions but 
different metallicities.  Brighter than the ``turn-over'' at $M_V \sim -7.5$, 
the luminosity functions (LFs) are well approximated by power-laws with 
an exponent of about $-1.75$.  This is similar to the LF for young star 
clusters, suggesting that young and old globular clusters form by the same 
basic mechanism.  We discuss scenarios for GC formation and conclude that 
our data appear to favor ``in-situ'' models in which all GCs in a galaxy 
formed after the main body of the proto-galaxy had assembled into a 
single potential well.
\end{abstract}

\keywords{galaxies:elliptical and lenticular,cD ---
          galaxies:evolution ---
          galaxies:star clusters}

\section{Introduction}

  Based on current observational evidence, swarms of globular clusters (GCs)
appear to surround virtually all large galaxies and quite a few lesser 
ones. In particular, it was noted early on that large elliptical galaxies 
like e.g.\ M87 are hosts to exceedingly rich GC populations \citep{baum55}. 
Before the advent of sensitive CCD detectors in the early 1980s, studies of
GCs in galaxies beyond the Local Group remained very demanding, but over
the past couple of decades an impressive amount of photometric data 
has been collected for globular clusters (GCs) in external galaxies.
Further progress has been made with data from the \emph{Hubble Space 
Telescope}, reducing contamination to a minimum because of its superior
resolution.

  Studies of extragalactic GCs have often aimed at characterizing the 
properties of all GCs around any given galaxy as a {\it system} and 
comparing these with the GC system of our own and other galaxies. Since GCs 
are generally 
thought to have formed very early, it has been anticipated that 
similarities and/or differences between GCSs in various galaxy types 
would eventually contribute to an improved understanding of the formation 
and early evolution of their host galaxies.

  Perhaps the most celebrated of the similarities between GCSs in
different galaxies is the apparently ``universal'' {\it luminosity 
function} of globular clusters. When plotted in magnitude units, the 
globular cluster luminosity function (GCLF) in practically all galaxies 
studied to date appears to be quite well fit by a Gaussian with a mean 
or ``turn-over'' at $M_V \sim -7.5$ and a dispersion of $\sigma_V\sim1.2$
\citep{har91,ash98}.  However, few studies have reached 
deeper than $\sim 1$ mag below the turn-over and there have been plenty 
of attempts to fit Gaussians to even shallower data. It is also 
worth noting that there is no {\it a priori} reason to prefer a Gaussian 
as a fitting function. In the few galaxies where the luminosity 
distribution of globular clusters is known to several magnitudes fainter than
the turn-over, other analytic functions actually provide a better fit.
\citet{sec92} found that in the Milky Way and M31, a $t_5$ function
is a significant improvement compared to a Gaussian. Although the
difference between a Gaussian and a $t_5$ function is relatively
minor when plotted in the standard magnitude space, it turns out to
be very conspicuous when the LFs are plotted in
{\it luminosity} units (Sect.~\ref{sec:lf_an}). 
  
  In recent years, HST observations have revealed extremely luminous 
{\it young} star clusters in starburst environments. For any reasonable 
M/L ratios, these young clusters have masses that are easily within the 
range spanned by typical old globular clusters, and it is tempting to 
assume that the study of such clusters may tell a great deal about how 
the old GCs formed. The luminosity functions of young clusters generally
follow power-law distributions of the form $n(L)dL \propto L^{\alpha}dL$.
For young clusters in the ``Antennae'', \citet{whi99} found an overall 
LF slope of $\alpha = -2.6$ above $\sim10^5\msun$ and $\alpha  = -1.7$ 
below this limit. However, \citet{zha99} found that the \emph{mass} 
function of Antennae clusters is well represented by a power law with 
exponent $-2$ over the entire mass range $10^4 < M < 10^6\msun$.  For young 
clusters in the recent merger NGC~3256, \citet{zep99} found a power-law 
with exponent $-1.8$ to be a good fit to the LF, and similar power-laws have 
been fitted to the LFs of massive young clusters in starburst galaxies 
like He 2--10 ($\alpha = -1.8\pm0.1$) and NGC~1741 ($\alpha = -1.9\pm0.1$) 
\citep{jo99,jo00}.  
For Milky Way open clusters \citep{van84} and young clusters in the LMC 
\citep{els85}, the LF is well represented by a power law with 
$\alpha \sim -1.5$.  However, when evolutionary effects are taken into account 
the slope of the \emph{mass} function may be closer to $-2$ \citep{elm97}. 
The mass/luminosity functions for a variety of other young objects 
(e.g.\ HII regions and Giant Molecular Clouds) are generally well represented 
by similar power-laws \citep{hp94,elm97}.

  Unfortunately, the fact that LFs of old globular clusters have
traditionally been discussed as a function of \emph{magnitude}, while
studies of young cluster systems often use linear luminosity units, makes
direct comparison of the LFs of young and old clusters more difficult.
This has sometimes led to the misconception that the LFs of young and old
clusters are dramatically different. However, the \emph{upper} part of 
the LF of old GCs and the LF of young clusters are, in fact, well 
represented by power-laws with very similar exponents.  For example, 
\citet{kis94} and \citet{kis96} found power-laws with a slope of 
$\alpha = -1.9\pm0.1$ to be a good fit to the mass function for old GCs 
brighter than the turn-over in the two ellipticals, NGC~4636 and NGC~720.
Important clues to the formation of old GCs may lie in the similarity of 
their LF to that of young objects rather than in the differences, which 
become apparent mainly for luminosities \emph{fainter than} the GCLF 
turn-over. Furthermore, little is actually known about the detailed 
behavior of the faint end of the GCLF, and in particular, whether the 
shape of the LF is really universal at the faintest levels in old as well 
as young cluster systems.


  Another important result concerns the \emph{color} distribution of old
GCs. One of the more striking recent discoveries has been that of {\it bimodal
color distributions} for GC systems \citep{els96,gei96,kis97}. This was 
originally predicted by \citet{ash92} as a consequence of 
gas-rich mergers. In their scenario, a merger product would contain a 
metal-poor (blue) GC population inherited from the progenitor galaxies and a 
metal-rich (red) population, formed in the merging process. After the 
initial excitement about the discovery of bimodal color distributions it 
has become clear that not all observed properties of GC systems fit into the
merger picture and alternatives have been suggested \citep{for97,cot98,hil99}.
It is probably fair to say that there is currently no consensus concerning 
theories for the origin of multiple globular cluster populations within 
galaxies \citep{rho01}, but bimodal color distributions have turned out to 
be very common \citep{geb99,kun99} and understanding them is bound to 
provide insight into processes that must have been common in the 
evolutionary history of galaxies.

  In spite of the often-quoted strong evidence in favor of the ``universal''
GCLF, there may be some reason to question its universality.  In a recent 
study of the nearby S0 galaxy NGC~1023, \citet{lar00} found a 
number of clusters with much larger effective radii (about 10 pc) than 
normal GCs. Interestingly, these clusters were also {\it fainter} and did 
not fit a ``standard'' GCLF.  It is also well-known that the outer, more
extended halo clusters in the Milky Way are generally fainter and do not
share the standard GCLF \citep{van83,van96}. This underscores the
need to explore the {\it faint} end of the GCLF and test whether it is
truly universal.  Combining the photometric data with information about 
the sizes of individual globular clusters may provide further clues to the 
origin of GC systems, e.g.\ in order to check whether the faint extended
clusters in NGC~1023 are a common phenomenon, or if they are unique to this 
galaxy and therefore presumably formed in some rare event.

  In the Milky Way, globular clusters typically have half-light (effective) 
radii of about 3 pc (Harris 1996) although clusters in the outer halo 
are significantly larger \citep{van96}. At the distance of the nearest
large galaxy clusters (Virgo and Fornax), this corresponds to about
$0\farcs04$, so whether these GC sizes
are typical in other galaxies remained completely unknown until the
HST era. Sizes of GCs beyond the Local Group have now been measured in a
number of galaxies and it appears that GCs in other galaxies have roughly
the same sizes as in the Milky Way. More surprisingly, it appears that
the {\it red} GCs are generally somewhat smaller than the {\it blue}
ones by 20 -- 30\% \citep{kun98,puz99,kuea99,lar00}. Whether this difference
was set up at formation or is due to dynamical evolution remains an
open question.

  The HST archive now contains images of more than 50 early-type galaxies 
and thus provides an invaluable database for studying and comparing GCSs 
around different galaxies. Such surveys have already been 
undertaken by other authors \citep{geb99,kun99} and we do not intend
to duplicate their efforts here. However, the previous studies
have put their main emphasis on a comparison of as many GCSs as
possible, not necessarily requiring homogeneity in depth and/or choice of
bandpasses. Here we aim at a more specific investigation of relatively 
nearby galaxies for which {\it deep} HST WFPC2 imaging is available in the
F555W and F814W bands, which can be accurately transformed to standard
$V$ and $I$ magnitudes.  Most of the data were obtained in Cycles 5 and 6 
by our group, where much of the Cycle 6 data is published for the first
time in this paper. We have supplemented our data with archive
data, but only data with exposure times comparable to those of our
Cycle 5/6 datasets have been included in order to reach well beyond the
GCLF turn-over and provide sufficient S/N for size measurements.
We have also obtained photometry for objects in two 
comparison fields as a check of background/foreground contamination.

\section{Analytic models of luminosity functions}
  \label{sec:lf_an}

  As shown by \citet{sec92}, the Student's $t_5$ function 
\begin{equation}
  \frac{8}{3 \sqrt{5} \pi \sigma_t} \, 
    \left(1 + \frac{(M-\mu_t)^2}{5 \sigma_t^2}\right)^{-3}
\end{equation}
  where $\mu_t$ is the turn-over magnitude and $\sigma_t$ is the 
dispersion of the $t_5$ function, provides a significantly better fit to 
the Milky Way and M31 GCLFs than a Gaussian.  This is illustrated in the 
left panel of Fig.~\ref{fig:gclf_mw} which
shows the Milky Way GCLF together with a Gaussian and a $t_5$ function,
both with a mean of $M_V = -7.3$ and dispersions of $\sigma_G = 1.4$ and
$\sigma_t = 1.1$ respectively \citep{sec92}. The $t_5$ function evidently 
provides a better fit than the Gaussian, particularly at the faint end. 
However, if one restricts the fit to clusters brighter than $1-2$ mag below 
the turn-over, as in most studies of extragalactic GCSs, the difference 
remains quite subtle. 

  The difference between the Gaussian and the $t_5$ function becomes much 
more clear when plotting the number of clusters per {\it luminosity} 
bin rather than per {\it magnitude} bin, as illustrated 
in the right-hand panel of Fig.~\ref{fig:gclf_mw}.  In luminosity units,
a log-normal distribution is actually quite far from providing a satisfactory 
fit below $\sim 10^4 \lsun$ or $M_V\simeq-5$, while the $t_5$ function 
does, in fact, match the observed luminosity function almost perfectly. 
Normalizing the two functions to the same value at the turn-over, the
difference between the $t_5$ function and the Gaussian amounts to
a factor of 4 at $M_V=-4$ and increases rapidly below this limit,
reaching a factor of 30 at $M_V=-3$.  So although the difference between
a Gaussian and a $t_5$ function is almost indiscernible for typical
extragalactic GC data, it becomes quite significant at faint levels.
This difference may turn out to be of great relevance to the understanding 
of dynamical destruction processes in GC systems, as a $t_5$--like 
distribution requires considerably fewer low-mass clusters to be destroyed 
for an initially uniform power-law mass distribution.

  In addition to Gaussian and $t_5$ functions, other analytical models
have been used to fit the mass- and luminosity functions of globular clusters 
as well. \citet{av95} used Gauss-Hermite expansions to characterize the
Galactic GCLF but found the higher-order terms to be small, indicating no
strong deviations from a pure Gaussian.  \citet{baum95} preferred 
a composite of two exponentials over Gaussian or $t$ functions as a fit 
to the combined LF of Milky Way and M31 globular clusters, corresponding 
to two power-law segments if luminosity units are used instead of magnitudes. 
A figure comparing the Gaussian, $t_5$ and exponential functions is given 
in that paper.  Two-component power-law fits were also used by 
\citet{hp94} to fit the mass distribution of old globular clusters, and
there is some evidence that the highest-mass clusters may follow
a third power-law mass distribution \citep{mac95}.



\section{Data}

  Basic data for the galaxies studied in this paper are listed in
Table~\ref{tab:gdata}.  The first column of the table gives the
galaxy name. Some galaxies have two or more pointings, usually with 
one pointing centered on the galaxy nucleus and another one offset from 
the center, denoted by a '-O' suffix.  The second column of 
Table~\ref{tab:gdata} gives the name of the original principal investigator
(PI) of the dataset, along with the ID (PID) of the proposal from which 
the data originate. Exposure times in sec in F555W and F814W are in 
cols.\ 3 and 4. A number of additional host galaxy parameters are given 
in Table~\ref{tab:props}, including infrared $JHK$ colors from the
2MASS survey \citep{jar00}.

  In three cases we have included data which are not in
the F555W/F814W bands: Two galaxies in the Fornax cluster (NGC~1399, 
NGC~1404) were observed in F450W / F814W and data for the ``Sombrero'' 
galaxy (M104) is in F547M / F814W. While the latter is easily transformed
to $\vi,V$ photometry, the NGC~1399 / NGC~1404 data is closer to the
$\bi,B$ bands and will only be transformed to $\vi,V$ (using transformations
in \citet{for00}) when this is essential for comparison with the other 
galaxies.  

  We have generally included galaxies for which integrations longer
than about 2000 sec were available in both F555W and F814W. However,
for a few nearby galaxies (NGC~3115, NGC~3379, NGC~3384 and NGC~4594) this 
requirement has been relaxed. 
In all cases, the exposures listed in 
Table~\ref{tab:gdata} actually consist of two or more shorter integrations.

  The data were downloaded from the archive at STScI and initial
reductions (flatfielding, bias subtraction etc.) were performed
``on the fly'' by the standard pipeline processing system.
Subsequent reductions were done with IRAF\footnote{IRAF is distributed 
by the National Optical Astronomical Observatories, which are operated by 
the Association of Universities for Research in Astronomy, Inc.~under 
contract with the National Science Foundation} and closely followed the
procedure described in \citet{lar00}. The individual exposures 
in each band were combined using the IMCOMBINE task, with the {\bf reject} 
option set to {\bf crreject} in order to eliminate cosmic ray (CR) hits.
In most cases, the individual exposures images were well aligned so that 
no shifts were required before combination, but when shifts were necessary 
they were applied with the IMSHIFT task.

  For the central pointings, a sky background subtraction was done.  First, 
point sources were subtracted from the raw images using the \ishape\ 
algorithm \citep{lar99}.  The object-subtracted images were then smoothed 
with a $15\times15$ pixels median filter and the smoothed images were then 
subtracted from the original images, providing our final set of 
background-subtracted images for further analysis. 

  Photometry was done with the PHOT package within IRAF. Point sources 
were detected using the DAOFIND task, but because of the varying S/N
within the images it was necessary to impose further selection
criteria in order to avoid spurious detections. This was done by
measuring the noise level directly off the images in a small annulus
around each object and only including objects which had a S/N higher than 
3 within an aperture radius of 2 pixels in both F555W and F814W.
Aperture photometry for all objects in the final object lists was
then obtained using the APPHOT task. For the background-subtracted images
a fixed background level of 0 was used, while the background was measured
between 30 and 50 pixels from the object in the off-center pointings.

The transformation from F555W / F814W magnitudes to the standard $V, \vi$ 
system was done following the standard procedure as described in \citet{hol95}.
Charge-transfer efficiency corrections from \citet{stet98} were also
applied, even though this correction is nearly negligible in our case 
because of the underlying galaxy light. We note that the recent
recalibration of WFPC2 data reduced with HSTphot \citep{dolp00} would 
cause \vi\ colors to be redder by 0.02 mag compared to the
Holtzman et al.\ calibration.

Colors were measured in an $r=2$ pixels aperture, but because globular 
clusters are expected to appear as slightly extended sources in our images 
we decided to use a slightly larger $r=3$ pixels aperture for $V$ magnitudes. 
For colors the use of a smaller aperture is justified even for extended 
objects because the \emph{difference} between aperture corrections in 
different bands is nearly independent of object
size \citep{hol96,lar00}. We have adopted aperture corrections from our
science aperture to the Holtzman et al. $0\farcs5$ reference apertures of
$\Delta V_{\rm WF} = -0.15$ mag and $\Delta (\vi)_{\rm WF} = 0.03$ mag for 
the WF chips and $\Delta V_{\rm PC} = -0.55$ mag and 
$\Delta (\vi)_{\rm PC} = 0.120$ mag for the PC chip, respectively. 
These aperture corrections are derived for objects that have King profiles 
with an effective radius of 3 pc and $r_{\rm tidal} / r_{\rm core} = 30$ 
at a distance of about 10 Mpc, typical for GCs in
galaxies such as NGC~1023 and NGC~3115. At the distance of Virgo 
($\sim16$ Mpc), GCs will have smaller apparent sizes and our magnitudes
may thus be too bright by a few times 0.01 mag. However, colors are not
affected.  For a thorough discussion of aperture corrections and their 
dependence on object size we refer to \citet{lar00}. 

  In addition to the data listed in Table~\ref{tab:gdata}, we 
included two comparison fields to check contamination by
foreground/background sources. One of these fields is located about
$2\deg$ from the galaxy NGC~1023 (From HST proposal 6254, PI: E. Groth). 
The data for this field consist of $2\times1300$ sec in each of the 
F606W and F814W bands, i.e.\ roughly as deep as our science data. The 
other field is the {\it Hubble Deep Field} (HDF) \citep{wil96} for which 
we combined F606W and F814W images to obtain total integration times of 
6300 sec and 12900 sec in the two bands, respectively. Although the 
transformation of F606W/F814W data to standard $V,\vi$ photometry is 
presumably less accurate than for F555W/F814W data, it should be good 
enough for our comparison purposes.

  Completeness tests were carried out by adding artificial objects to
the science images and redoing the photometry. 500 artificial objects
were added to each chip at random positions, but with required
separations larger than 10 pixels in order to avoid crowding
problems. The artificial objects were added to the science images using 
the \mksynth\ task \citep{lar99}, again generating the object profiles by
convolution of King profiles similar to those used for the aperture
corrections with the WFPC2 PSF as 
modeled by the \tinytim\ program \citep{kri97}. Here the King profiles
were scaled according to the distances listed in Table~\ref{tab:props}.
In principle, the completeness 
corrections will be color dependent, but GCs generally span a relatively 
narrow range in colors (typically between $\vio = 0.8$ and 1.2) so for these 
tests we assumed constant object colors of $\vi=1.0$, which is close to the 
average colors of GCs. The 50\% completeness limit was
generally found to be between $V=25$ and $V=26$, in most cases well below 
the expected turn-over of the globular cluster luminosity functions.

\section{Description of galaxies}

  Our sample consists of 17 galaxies of which 1 is classified as
type Sa, 4 are classified as S0, 11 are ellipticals and 1 is a cD.
In the following we give a short description of each galaxy with
references to earlier work pertaining to their GC systems.

\noindent
{\bf NGC 524}\\
This massive S0 galaxy dominates a small group. It has a relatively rich GC 
system with a specific frequency S$_N$ \citep{hv81} $\sim$ 3.3 
\citep{ash98}. Here 
we detect a total of 617 GCs, second only to the cD galaxy NGC 4486 (M87).

\noindent
{\bf NGC 1023}\\
At 9.8 Mpc, NGC 1023 is the nearest S0 galaxy. It is the brightest galaxy 
in a group of 13 galaxies \citep{tul80}. Its GC system  has been studied 
recently using HST by \citet{lar00}. As well as the expected blue and red 
GC subpopulations, they identified a third subpopulation of red, spatially 
extended clusters. The origin of these extended clusters is currently 
unknown. 

\noindent
{\bf NGC 3115}\\
This bulge--dominated S0 galaxy is also very nearby. It has a modest GC 
system, which has been examined using HST by \citet{kun98}.
They identified 144 GCs with a similar color distribution to that listed 
in Table~\ref{tab:kmm}.  In addition they noted that the red GCs were 
smaller than the 
blue ones by $\sim$20\%. Mainly because of our selection criteria, designed
for somewhat more distant galaxies than NGC~3115 and therefore with a
brighter lower magnitude limit, we detect fewer GCs in NGC~3115
than \citet{kun98}.

\noindent
{\bf NGC 3379}\\
Also known as M105, NGC 3379 is the dominant elliptical in the 
nearby Leo group. Along with NGC 3377, it has a relatively low specific 
frequency, i.e. S$_N$ $\sim$ 1.2 \citep{ash98}. Here we detect 
only 55 GCs with bimodality likely, but not certain.

\noindent
{\bf NGC 3384}\\
Also in the Leo group, this S0 galaxy has S$_N$ $\sim$ 0.9 \citep{ash98}. 
We detect a total of 54 GCs, of which 30 formally fall in the ``blue''
category and 24 in the red, but with no statistical evidence for bimodality.

\noindent
{\bf NGC 4365}\\
Probably located slightly behind the Virgo cluster, this large elliptical 
contains a kinematically distinct core \citep{ben88} which may indicate a past 
merger event. The first HST study was that of \citet{forb96}. From a 
short exposure, central pointing they detected 328 GCs but no obvious 
bimodality. Here we use longer exposure images of the galaxy center and a 
region directly to the north. We find a broad color distribution for 323
clusters brighter than $V=24$, although most GCs seem to belong to a peak 
with the same color as the blue population in most large ellipticals.

\noindent
{\bf NGC 4406}\\
This luminous early-type (E3/S0) Virgo galaxy is also known as M86. It
is a rare example of an early type galaxy with an X-ray plume, presumably 
resulting from ram pressure stripping \citep{ran95}.  A 
kinematically distinct core was found by \citet{ben88}. It was part of the 
same HST study by \citet{forb96} which included NGC~4365. Like 
NGC~4365, no obvious bimodality was detected in the short exposure images.  
\citet{kun99} found tentative evidence for bimodality with peaks at 
\vio = 0.98 and 1.17; here we formally find peaks at $\vio = 0.986$ and 1.145 
but still with bimodality detected at a low confidence level.

\noindent
{\bf NGC 4472}\\
Also known as M49, this galaxy is the most luminous in the Virgo cluster,
dominating its region of the subcluster. It has a joint elliptical and S0 
classification and a kinematically distinct core \citep{dav88}.
NGC 4472 has been the subject of numerous GC studies and contains perhaps the 
best characterized GC system of any early--type galaxy.  Spectroscopic 
studies \citep{bri97,bea00,zep00} have determined the mean metallicity of 
the GC system and begun to constrain the age and kinematics.  Photometric 
studies \citep{lee98,puz99,lee00,rho01} have detected large numbers of GCs
with clear evidence for bimodality.  In terms of the peak colors, there 
are slight differences between our results and the other HST studies even 
though all three use essentially the same data set. We find $\vio$ = 0.94 
and 1.21, which is consistent with the findings of \citet{lee00} and 
\citet{puz99} after extinction corrections. We also confirm that the
turn-over magnitude of the red GCs is fainter than for the blue GCs, as
noted by \citet{puz99}.

\noindent
{\bf NGC 4473}\\
This is a highly elongated Virgo elliptical. \citet{kun99} reports bimodality 
from HST data. Our detection of bimodality is statistically not compelling, 
but we find similar mean color peaks to Kundu.

\noindent
{\bf NGC 4486}\\
As the Virgo central elliptical galaxy (M87) it lies at the center 
of the Virgo cluster 
gravitational potential and X--ray emission, although it is moving slightly
with respect to the cluster center of mass velocity as determined from galaxy
velocities \citep{huc85}.  Its GC system is exceptionally 
rich with over 10,000 GCs and a S$_N$ $\sim$ 14 \citep{ash98}.  HST 
observations have allowed the bimodality to be well defined \citep{els96}.
Spectroscopic studies have derived metallicities and kinematics for a 
large number of M87 GCs \citep{mou87,hb87,coh98}.

\noindent
{\bf NGC 4494}\\
Located in the Coma I cloud, this elliptical contains a
kinematically distinct core \citep{ben88}. From HST 
data, \citet{forb96} suggested a low specific frequency of S$_N$ $\sim$ 2.
The same HST data suggested mean color peaks around $\vio = 0.95$ and 
1.15, here we find 0.92 and 1.12.

\noindent
{\bf NGC 4552}\\
This Virgo elliptical is also known as M89, and is known to posses a 
kinematically distinct core \citep{sim97}.  \citet{kun99} found possible 
bimodality, which we confirm as statistically significant with the same 
mean colors (to within $\pm$ 0.01).

\noindent
{\bf NGC 4594}\\
Known as `The Sombrero' (M104), it is the closest Sa type galaxy. With 
an exceptionally large bulge/disk ratio of $\sim6$ \citep{ken88}, it
represents an intermediate case between ellipticals and early type 
spirals. Its GC system is perhaps the most populous system of any
spiral galaxy with 1200 $\pm$ 100 estimated by \citet{har84}.
Photometry \citep{forb97} and spectroscopy \citep{bri97} of the GC 
system indicates a mean metallicity similar to that for elliptical 
galaxies. Recently, \citet{lar01} have utilized three HST pointings
of M104 to conduct a detailed study of the GC system. They detected 
strong color bimodality, and found the red GCs to be $\sim$30\% smaller 
than the blue ones. The Sombrero data used in this study are the same
as those used by \citet{lar01}.

\noindent
{\bf NGC 4649}\\
Another giant Virgo elliptical (known as M60), with S$_N$ $\sim$ 6.7 
\citep{ash98}. Both ourselves and \citet{kun99} detect clear bimodality.

\noindent
{\bf NGC 4733}\\
This galaxy is a low luminosity Virgo elliptical. We detect only 28 GC
candidates, too few to draw conclusions about bimodality although most
clusters appear to belong to a blue peak.

\noindent
{\bf NGC 1399}\\
Although of rather modest optical luminosity, NGC 1399 is the central
elliptical galaxy of the Fornax cluster. It has around 5000 GCs giving it 
a high specific frequency \citep{bri91}. Studies of the kinematics of the
NGC~1399 GC system have been carried out by \citet{gril94} and \citet{kis98}.
Here we use the same $B$ and $I$ band HST data of \citet{gril99}, but 
convert the color peaks to \vi .

\noindent
{\bf NGC 1404}\\
This galaxy lies within the X--ray envelope of NGC 1399, and may be losing 
GCs to it. Also part of the \citet{forb98} and \citet{gril99} photometric 
studies, we confirm a bimodal color distribution with about the same peaks as 
for the GCs in NGC~1399.

\section{Results}

\subsection{Color-magnitude diagrams}

  Color-magnitude diagrams (CMDs) for objects in the 17 galaxies in our
sample are shown in Fig.~\ref{fig:cmd}. Typical errors in $\vi$ are
indicated by the error bars at $\vio=2.0$. Note that the CMDs for NGC~1399 
and NGC~1404 are in $\bio,B$ units, although the magnitude and color
ranges shown for these galaxies have been scaled to match those of the
$\vio,V$ plots.  The horizontal dashed lines indicate approximate 50\% 
completeness limits from the completeness tests. 
As mentioned above, these completeness limits 
are for objects with typical GC sizes and colors; objects with larger sizes
will have brighter 50\% completeness limits. Note, however, that we 
reach well below the expected GCLF turn-over at $M_V\simeq-7.5$ 
($M_B\simeq-6.8$) in all galaxies.

  The GC sequences are easily recognizable in Fig.~\ref{fig:cmd},
extending over nearly the entire plotted magnitude range and with colors 
between $\vio\simeq0.8$ and $\vio\simeq1.25$ (the blue objects in the NGC~4649 
field with $\vio<0.5$ actually belong to the nearby spiral galaxy NGC~4647, 
located $2\farcm5$ from NGC~4649).  Although some CMDs exhibit strikingly 
bimodal color distributions (e.g.\ NGC~1023, NGC~1404, NGC~4472 
and NGC~4649) and others show less evidence for two distinct peaks in the 
\vi\ colors, the {\it total range} in \vio\ spanned by the GC sequence is
always much larger than the photometric errors.  For further analysis, 
globular cluster (GC) candidates were selected within the color range 
$0.70 < \vio < 1.45$ ($-2.2 < \feh < 0.2$)
and individually adjusted magnitude limits, as 
indicated by the boxes superimposed on the CMDs in Fig.~\ref{fig:cmd}.

  Fig.~\ref{fig:cmd_cmp} shows the CMDs for the two comparison fields.
The HDF appears to contain more objects than the other comparison field, 
perhaps partly as a consequence of the much longer exposures which make 
it easier to detect extended sources. Comparing Fig.~\ref{fig:cmd_cmp} and
Fig.~\ref{fig:cmd}, we see that contamination is unlikely to pose much
of a problem for the richer GCSs. However, for the poorer systems one
obviously needs to carefully address the contamination issue.

\subsection{Color distributions and bimodality}

In order to make more quantitative statements about bimodality, a KMM test 
\citep{ash94} was applied to the data. To reduce photometric errors, only 
clusters brighter than 1.0 mag above the lower magnitude limit 
indicated by the boxes in Fig.~\ref{fig:cmd} were included.
The KMM test uses a maximum-likelihood technique to estimate the probability 
that the distribution of a number of data values (in this case the \vi\ 
colors of GCs) is better modeled as a sum of two Gaussians than as a single 
Gaussian, as indicated by the number $1-P$. Here 
we have used a homo-scedastic test, i.e.\ the two Gaussians are assumed to 
have the same dispersion. The KMM algorithm also supports heteroscedastic
fits (different dispersions) but these tend to be more unstable and are
generally not recommended unless there are strong reasons to expect that
the dispersions are significantly different \citep{ash94}.

  Table~\ref{tab:kmm} lists the colors of the two peaks, the $P$ value, and 
the numbers of clusters assigned to each peak by KMM.  Fig.~\ref{fig:kmm} 
shows histograms for the \vio\ colors of the GCs fitted by the KMM test, 
together with Gaussians corresponding to the two color peaks and their sum.
In spite of the visual impression that many of the color distributions 
may not be convincingly bimodal, the $P$ value is actually close to 
0 in nearly all cases, indicating a high probability that two 
Gaussians are a better fit than a single one. Furthermore, the KMM test
returns consistent \vio\ peak colors even in galaxies where the
CMDs and \vio\ histograms show weak or no evidence of bimodality.
However, we note that NGC~4365 does appear to have only one peak,
centered on $\vio = 0.98$. Thus, the GCs in this galaxy have colors
typical for the blue (metal-poor) populations in the bimodal systems,
while the red GC population seems to be largely absent. 

  The average colors of the blue and red peaks are $\vio=0.95$ and
$\vio=1.18$, with a scatter of about 0.02 mag and 0.04 mag for the two
peaks, respectively.  The \vio\ colors can be converted into metallicities 
e.g.\ using the calibration in \citet{kis98}, yielding $\feh = -1.4$ and 
$\feh = -0.6$. These metallicities are roughly similar to those of the 
metal-poor (halo) and metal-rich (disk/bulge) clusters in the Milky Way 
\citep{zin85} and in M31 \citep{barm00}.
The relation given in \citet{kun98} leads to almost exactly
the same metallicity for the blue peak and a somewhat higher
metallicity of $\feh = -0.3$ for the red peak.

  Of course, there is no {\it a priori} basis for the assumption that
Gaussian functions are the best possible representation of the data.
One may even dispute that a low $P$ value is necessarily an indicator 
of bimodality \emph{per se}, since any observed distribution that is 
broadened relative to a Gaussian distribution will generally result in 
a low $P$ value. An alternative test for bimodality is the so-called 
``DIP statistic'' \citep{hh85,geb99} which measures the probability that 
a distribution is not unimodal, without any underlying assumptions about
the details of the distribution. DIP values are listed in the last column 
of Table~\ref{tab:kmm}. The DIP probabilities are generally high for 
those color distributions that also visually appear to be strongly bimodal
(e.g.\ NGC~1023, NGC~4649, NGC~4472, NGC~4486), and conversely, galaxies like 
NGC~524 and NGC~4365 have low DIP values.  

\subsection{Turn-over magnitudes}

  Several recent studies have found significant differences between
the turn-over magnitudes for the LFs of blue and red GCs, with the blue GCs
generally being brighter by a few tenths of a magnitude in the $V$ band
\citep{els96,kuea99,puz99,lar00}. For a constant globular cluster
\emph{mass} distribution and similar ages, a difference in turn-over 
magnitudes is an expected consequence of the variation in mass-to-light 
ratio with metallicity.  For two equally old GC populations with 
metallicities of $\feh = -1.4$ and $\feh = -0.6$ and similar stellar IMFs,
\citet{acz95} find that the difference in $V$ band magnitude amounts to 
about 0.22 mag.  We also estimated mass to light ratios for simple stellar 
populations with $\feh = -1.4$ and $\feh = -0.6$ using 1996 versions of 
the Bruzual \& Charlot population synthesis models and found an expected 
difference of 0.26 mag in the $V$-band turn-over magnitudes for two 
populations of similar old ages (12 -- 15 Gyr).  It should be noted that 
the dependence of mass-to-light ratio on metallicity is wavelength 
dependent and becomes weaker at longer wavelengths.  

  In the following we will refer to ``blue'' and ``red'' GC candidates as
clusters with $\vio<1.05$ and $\vio\ge1.05$, respectively.  The \vio\ cut 
corresponds to a metallicity of $\feh = -1.1$ \citep{kis98} and is close
to the natural division between halo and disk/bulge GCs in the
Milky Way \citep{zin85}.  We performed maximum-likelihood fits of $t_5$ 
functions \citep{sec92} to the luminosity distributions of our blue and 
red GC candidates, selected within the boxes in Fig.~\ref{fig:cmd}.
Completeness corrections were done on a 
by-chip basis and a correction for contamination was performed by a 
statistical subtraction of objects in the NGC~1023 comparison field from 
the source lists. As a check of the maximum-likelihood fits, we also 
fitted Gaussian functions directly to histograms of the raw luminosity 
functions in a few cases, using the NGAUSSFIT task in the STSDAS package. 
The GCLF turn-over peaks estimated by the two methods typically agreed 
within about 0.15 mag, which is quite satisfactory considering that the 
Gaussian fits were performed on data that were corrected neither for 
completeness effects nor contamination.

  To test how different completeness functions for blue and red GCs might
affect the turn-over magnitudes, we carried out additional completeness 
tests for objects with $\vi = 0.8$ and $\vi = 1.2$ in two fields (one 
off-center pointing in NGC~4472 and the central pointing in NGC~4486).  The 
$t_5$ function fits were then repeated for all combinations of GC 
subpopulations and completeness functions.  Being at the extremes of the 
GC color distribution, these tests provide a worst-case estimate of how 
much color-dependent completeness corrections might affect the turn-over 
differences between blue and red GC populations.  Regardless of the choice 
of completeness functions, the turn-over magnitudes remained constant to 
within 0.05 mag. 

  The results of our maximum-likelihood fits are listed in 
Tables~\ref{tab:mto_var} and \ref{tab:mto_fix} for blue and red 
subpopulations separately, as well as for 
the combined samples.  The last column of each table lists the difference 
\dmto\ between the turn-over magnitudes of the blue and red GC populations 
in each galaxy.  In Table~\ref{tab:mto_var}, both the dispersion
($\sigma_t$) and turn-over were fitted, while the dispersion was kept
fixed at $\sigma_t = 1.1$ in Table~\ref{tab:mto_fix}. This dispersion
corresponds to the value reported for the Milky Way GCS by \citet{sec92}.

  Indeed, we find that the blue GCs are generally brighter than the reds 
by typically a few times 0.1 mag. The difference is significant in all 
galaxies, including those without obviously bimodal color distributions 
(e.g.\ NGC~524, NGC~4365). In particular, we confirm the offset between 
the turn-over magnitudes of the two GC populations in NGC~4472 which was 
first reported by \citet{puz99}, but subsequently claimed not to exist 
by \citet{lee00} based on essentially the same data. 
  Averaging all the \dmto\ values in Table~\ref{tab:mto_var} and weighting
each value inversely by the sum of its positive and negative errors yields a
mean difference between the turn-over of the red and blue GC populations
of 0.47 mag.  Within the uncertainties, all \dmto\ values in the table are 
consistent with this number, which is somewhat larger than the prediction 
by population synthesis models. It is perhaps worth noting that a few galaxies
show much larger \dmto\ values than the typical $\sim 0.5$ mag, even though
the formal errors on \dmto\ are also larger than average in those cases.
In particular, the two galaxies NGC~1023 and NGC~3384 both have 
$\dmto > 1$. These two galaxies both appear to contain a third population
of clusters which are intrinsically fainter than `real' globular clusters
and predominantly red \citep[][see also Sect.~\ref{sec:lf} and 
\ref{sec:gcsize}]{lar00} and
may be responsible for shifting the fit for the red GCs towards fainter
magnitudes.  If NGC~1023 and NGC~3384 are excluded the average $\dmto$ 
value decreases to 0.40 mag with a scatter of $\pm0.24$ mag.

  If we use the fixed-$\sigma_t$ fits in Table~\ref{tab:mto_fix} instead of
the two-parameter fits in Table~\ref{tab:mto_var} and again exclude
NGC~1023 and NGC~3384, the average \dmto\ value decreases slightly to 
0.37 mag with 
a scatter of $\pm0.20$ mag. Note that the \dmto\ values for NGC~4472 and 
NGC~4365, which were in both cases larger than 0.80 mag for the 
two-parameter fits, now decrease to 0.50 mag and 0.56 mag, respectively.  
However, for NGC~1023 and NGC~3384 the \dmto\ values remain unusually 
large also for the one-parameter fits.

  Although one may potentially obtain a better fit to the data by letting
the dispersion vary, the two-parameter fits are also more sensitive to
outlying data points, inaccurate contamination and/or completeness
corrections etc. It is therefore not surprising that the scatter in \dmto\
decreases
when only one parameter (the turn-over magnitude) is fitted, and it 
seems reasonable to assume that the one-parameter fits yield somewhat more 
accurate estimates of the turn-over magnitudes, especially for the
cluster-poor GC systems.  This should certainly be the case if the GCLF 
is truly universal, with a (nearly) constant dispersion from galaxy to 
galaxy.  If the sample is further 
restricted to galaxies for which the errors (average of positive and negative) 
on \dmto\ are $<0.25$ mag, then we get an average $\dmto = 0.30$ for both 
one- and two-parameter fits, with a scatter of $\pm 0.16$ mag in both cases.
Of course, selecting the galaxies by the errors on the $t_5$ fits introduces 
a bias towards cluster-rich systems, which might have systematically 
different properties from the poorer GC systems, so whether or not the 
decrease in \dmto\ for the error-selected sample is real is hard to tell.
In any case, the average \dmto\ values are quite close to the theoretical 
expectation for two GC populations of similar ages and mass distributions, 
but different metallicities.

  In Table~\ref{tab:mto_fix} we have also included $t_5$ function fits
to Milky Way globular clusters (from Harris 1996). Following \citet{sec92}, 
we have only included clusters between 2 kpc and 35 kpc from the Galactic 
center and with color excess $E(\bv) < 1.0$. We exclude clusters fainter 
than $M_V=-5$, roughly corresponding to the magnitude limit in the other 
galaxies in our sample. These selection criteria leave only 67 ``blue'' 
($\feh < -1$) and 20 ``red'' ($\feh \ge -1$) clusters, so we decided not 
to attempt two-parameter fits for these data. From one-parameter fits to 
the Milky Way data we find that the turn-over of the blue GCs is
brighter than that of the red ones by $\dmto = 0.46^{+0.34}_{-0.38}$ mag, 
which is similar to the average \dmto\ value for the other galaxies in the
sample within the uncertainties.  It is also of interest to compare
with the GC system of M31: Curiously, \citet{barm01} found that 
the metal-poor (blue) globular clusters in M31 are on average 0.36 
\emph{fainter} in $V$ than the metal-rich (red) ones, a difference of about 
the same magnitude but opposite sign compared to most other GC systems.
\citet{barm01} argue against any possible selection effects as the cause 
of the measured GCLF differences in their sample, but note that
independent confirmation using a more complete and less contaminated
M31 cluster catalog would be highly desirable.

  However, it is worth reiterating that a comparison of the GCLF
turn-over magnitudes for different GC populations only makes sense for 
strictly identical \emph{mass} distributions of the various populations.
Establishing the mass distributions of GCs in external galaxies independently 
of the luminosities will be a very difficult task observationally.  If the 
two populations did, in fact, form at different epochs in different 
environmental conditions, then one might indeed \emph{expect} their mass 
functions to be different -- especially considering that the mass function 
of clusters in present-day starbursts deviates from that of old GCs in the 
critical region near and below the GCLF turn-over. We therefore feel that 
it would be premature to draw further conclusions about age differences 
between GC subpopulations, based only on differences in the GCLF turn-over.



\subsection{Luminosity functions}
\label{sec:lf}

  In Fig.~\ref{fig:ldf} we show the luminosity functions for GCs in each 
of the 15 galaxies with $V,I$ photometry.  We use luminosity units rather 
than magnitudes; for this 
reason the familiar ``Gaussian'' shape of the GCLF is not apparent in the 
figure.  The GCLF turn-over magnitude of $M_V = -7.5$ corresponds to about 
$8\times10^4 \, \lsun$.  For comparison, we have also included the LF
for Milky Way globular clusters (shown with dots), using data from 
\citet{har96}.
The solid lines represent the raw LFs of GCs in each galaxy, 
uncorrected for completeness and/or contamination effects, while the two 
dashed lines in each plot represent the LF corrected for incompleteness 
and background 
contamination, using each of the two reference fields.  For the HDF, the 
contamination correction was performed \emph{after} completeness correction, 
since the HDF data are much deeper than any of our GC data. For the other 
comparison field we applied the contamination correction \emph{before} 
correction for incompleteness.
The HDF is somewhat richer in background galaxies than the field near 
NGC~1023, while the latter is located at lower galactic latitude ($b=19\deg$) 
and thus presumably contains more Galactic foreground stars. The difference 
between the two dashed curves is likely to provide a rough indicator of the 
accuracy by which the LFs can be determined.  The average 50\% completeness 
limits are indicated by vertical dashed lines.

  In order to compare with the LFs of young cluster systems, we performed 
power-law fits to the LFs of our GC data in the interval 
$10^5 < L < 10^6 \, \lsun$ for red and blue clusters separately, and for 
the combined sample (Table~\ref{tab:ldf_up}). In contrast to $t_5$ function 
or Gaussian fits where both the dispersion and mean can vary, 
power-law fits to the brighter portion of the LF are completely independent of 
the behavior of the data below the turn-over and therefore hardly affected
by completeness effects.  The numbers in Table~\ref{tab:ldf_up} are for the 
raw data, i.e.\ not corrected for completeness or contamination. 
Applying these (uncertain) corrections changes the exponents by at most 
a few times 0.01. A weighted mean of data with errors on the power-law
exponents less than 0.5 gives exponents of $-1.66\pm0.03$ and $-1.80\pm0.06$
for the blue and red clusters, respectively, and $-1.74\pm0.04$ for the 
combined sample. These values are in very good agreement with those
reported for a variety of \emph{young} cluster systems (see the Introduction).
The slight difference between the fits to the red and blue clusters is 
most likely due to the differences in the turn-over magnitudes of the 
two populations. 

  There are a few galaxies in which the \emph{faint} end of the LF deviates 
significantly from that observed in the Milky Way. The most striking example 
is NGC~1023 which exhibits a strong excess of faint clusters \citep{lar00}, 
readily visible by inspection of the color-magnitude diagram
(Fig.~\ref{fig:cmd}), where especially the red GC sequence is seen to extend 
to very faint magnitudes. A similar effect is seen in NGC~3384,
and also the Sombrero, NGC~4472, NGC~4365 and NGC~4649 may show a hint of 
this phenomenon although incompleteness and contamination problems at the 
faint end of the GCLF make the results less conclusive for these galaxies. 
Nevertheless, there
might be reason to suspect that the GCLF is not as universal as previously
thought and it would be desirable to have even deeper HST data for some 
of these galaxies.  Variations from galaxy to galaxy at the faint end of 
the GCLF would have strong implications for theories for the formation and 
evolution of GC systems.  

\subsection{Sizes}
\label{sec:sizes}

  The spatial resolving power of HST allows sizes of extragalactic GCs
to be measured.  Some recent studies have shown 
a general trend for the red clusters to be somewhat smaller than the 
blue ones, both in elliptical galaxies like NGC~4472 and M87 
\citep{puz99,kun99}, in S0 galaxies such as NGC~3115 and NGC~1023 
\citep{lar00,kun98} and even in the Sa-type galaxy M104, the ``Sombrero'' 
\citep{lar01}.

  Here we discuss in some detail the size distributions of GCs in our
sample of galaxies.  Because of the undersampling of the point spread 
function (PSF) by especially the WF cameras, it is necessary to take special
care in avoiding instrumental effects. Here we have used the \ishape\
algorithm \citep{lar99} which models the cluster images as an analytic 
model profile convolved with the HST point-spread function, iteratively 
adjusting the model until the best possible match with the data is
obtained. We have chosen King profiles \citep{king62} with a concentration 
parameter of $c=30$ 
(tidal / core radius) for the analytic models.  Since the total integrated 
luminosity of King profiles is finite, the FWHM values measured by 
\ishape\ can easily be converted to half-light (effective) radii (\reff ).
All internal computations by the algorithm are performed on 10$\times$ 
subsampled image arrays, except for a final convolution with the so-called 
``diffusion kernel'' \citep{kri97}. For the size measurements we use
our F555W images, for which the diffusion kernel is best understood.

  For the two off-center pointings in NGC~4472 and the central pointing 
in NGC~4365, the individual exposures had to be shifted before combination, 
inevitably altering the PSF and potentially causing systematic errors in 
the size measurements. In these cases, sizes were therefore measured 
separately on each of the raw exposures
and then averaged.  This also allowed us to obtain an estimate of 
the accuracy of the size measurements, by computing the rms deviation of 
the difference between sizes of clusters measured on the two sets of images.
This test indicates an rms scatter of about 1.0 pc for the cluster sizes
measured on individual exposures (assuming Virgo distance) down to
$V=23.5$. Sizes measured on combined images (either directly or by
averaging two measurements) should therefore be accurate to about 1.0 pc
down to $V=24$ and in the following we adopt this magnitude limit for 
the size measurements.

  Possible systematic effects in the size measurements, resulting from 
the choice of a particular fitting function, are discussed in \citet{lar99}.
Briefly, the \emph{effective} radius is quite independent
of the choice of fitting function as long as the sources have comparable
sizes to the PSF (as in our case).


\subsubsection{GC size versus color}
\label{sec:gcsize}

  Fig.~\ref{fig:visz} shows the GC sizes as a function of \vio\ color
(\bio\ color for NGC~1399 and NGC~1404) while Fig.~\ref{fig:szdist} shows 
histograms for the size distributions of red and blue clusters separately.
The median sizes are listed in Table~\ref{tab:sizes}, which also gives 
cluster sizes in four narrower bins.  In order to minimize the effect 
of ``outliers'' with 
abnormally large sizes, the median sizes in Table~\ref{tab:sizes} are based 
on clusters with $\reff < 10$ pc only. For comparison, Table~\ref{tab:sizes} 
also includes data for GCs in the Milky Way \citep{har96}. For the Milky 
Way GCs, the division between ``red'' and ``blue'' clusters has been 
taken to be at \feh =$-1$. 

  Figs.~\ref{fig:visz} and \ref{fig:szdist} confirm that, in most cases, 
blue clusters are larger than red ones by typically $\sim20\%$. In
particular, this is true also for the two spiral galaxies in the table, 
i.e.\ the Sombrero and the \emph{Milky Way}. There are a few notable 
exceptions where blue and red clusters appear to have similar sizes, 
including NGC~4365 and NGC~4552, although the size difference is often
recovered when the reddest sub-bin ($1.20<\vio<1.45$) is excluded.

  As discussed by \citet{lar00}, the S0 galaxy NGC~1023 contains a 
population of extended, red clusters which are generally fainter than the 
``normal'' compact, globular clusters. These ``faint fuzzies'' are clearly 
visible in Fig.~\ref{fig:visz}, with sizes around 10 pc and a \vio\ color 
of about 1.2. They are responsible for much of the excess of faint clusters 
relative to the Milky Way GCLF seen in Fig.~\ref{fig:ldf} \citep{lar00}. 
Among the remaining galaxies in our sample, a similar population of extended 
red objects appears to exist in
NGC~3384. Apart from the generally poorer GCS of this galaxy, the
appearance of its $\vio, \reff$ diagram mimics that of NGC~1023. Furthermore,
as mentioned in Sect.~\ref{sec:lf}, the GCLF of NGC~3384 shows an excess
of faint clusters very reminiscent of that in NGC~1023. We thus suggest
that NGC~3384 is the \emph{second case} of a galaxy which hosts a
population of these, hitherto unknown, extended red faint clusters.

  Unfortunately, since the ``faint fuzzies'' are generally fainter than 
the GCLF turn-over, they are at the limit of reliable size measurements in 
the more distant galaxies in our sample. Furthermore, their extended nature 
and low surface brightness make their very detection more difficult than 
for ``normal'' compact GCs.  It is therefore difficult to tell how common 
such objects are and deep HST imaging of more nearby galaxies would be 
highly desirable.  Of the four galaxies in our sample where the faint 
extended clusters are readily detectable (NGC~1023, NGC~3115, NGC~3379 and 
NGC~3384), they are apparent in two and not detected in the other two. 
In particular in NGC~3115, with its relatively rich GCS and adequately 
deep photometry, there is little chance that a population of extended 
red clusters could have been overlooked. 


  In Fig.~\ref{fig:szplot}, the median GC sizes are plotted for each of the
four \vi\ sub-bins in Table~\ref{tab:sizes}
with symbol sizes proportional to the logarithm of the number of clusters. 
For easier comparison, cluster sizes have been normalized to 1.0 in the 
$0.90<\vio<1.05$ bin (which typically contains the largest number of GCs). 
Note that the scatter in the GC sizes is much larger in the reddest bin.
Interestingly, the cluster sizes decrease as a function of \vi\ for the 
three bluest bins, but then tend to increase somewhat again in the reddest bin. 
In some galaxies like NGC~1023 and NGC~3384, this effect is possibly due 
to the ``faint extended'' clusters which have preferentially red colors
while the cause is less obvious in other galaxies.

\subsection{Trends with galactocentric distance}

  Correlations between cluster properties and distance from the center of
their host galaxies can potentially provide further clues to the formation
and evolution of the cluster systems. In this respect, HST studies are
generally limited by the relatively small field of view: at the distance
of Virgo, a WFPC2 exposure centered on the galaxy nucleus will reach
out to about 8 kpc. For more distant galaxies the coverage obviously 
improves, but with the trade-off of decreased ability to measure cluster 
sizes.  Here we will carry out a ``case study'' of three cluster-rich 
galaxies (NGC~4365, NGC~4472 and NGC~4486) in the Virgo cluster, all 
with more than one WFPC2 pointing.

  Table~\ref{tab:sz_r} lists the sizes of blue and red clusters, colors 
of the blue and red peaks and numbers of blue and red clusters as a 
function of galactocentric distance $R_g$ for these three galaxies. Here 
clusters are defined as ``blue'' or ``red'' with respect to the color cut 
at $\vio=1.05$.  Sizes for individual clusters and peak colors were 
measured in the same way as described in the previous sections.

  None of the galaxies shows any significant color gradients for either
the blue or red peak. However, when plotting the ratio of blue to red 
clusters $N_{\rm blue} / N_{\rm red}$ as a function of $R_g$, both NGC~4486
and NGC~4472 show an increase in this ratio outwards (Fig.~\ref{fig:rplot}). 
In NGC~4486 the
increase in $N_{\rm blue} / N_{\rm red}$ is quite dramatic and almost certainly
responsible for the color gradient reported by earlier studies
\citep{stro81,lee93,coh98,har98}. A similar conclusion was reached by
\citet{kuea99} for the central parts of NGC~4486.
In NGC~4472 the trend in $N_{\rm blue} / N_{\rm red}$ vs.\ $R_g$ is more 
subtle and mostly driven by the outermost bin, in good agreement with the 
analysis by \citet{puz99} who concluded that the relative numbers of blue 
and red clusters in NGC~4472 change by no more than about 10\% within the 
central $250\arcsec$. However, further out in the halo NGC~4472 shows
a stronger decline in the number of red GCs \citep{gei96}.
NGC~4365 actually shows a \emph{decrease} in the
$N_{\rm blue} / N_{\rm red}$ ratio outwards, although a slope of 0 is
nearly within the error bars. It thus appears that the colors (and, by
inference, metallicities) of GC subpopulations are largely \emph{independent} 
of distance from the galaxy center. Any overall gradients seem to result 
only from a change in the mix of the subpopulations.

  Concerning overall size trends as a function of $R_g$, only NGC~4365 
shows a clear increase in the cluster sizes in the outermost regions. Apart 
from this, the GC sizes appear to be fairly constant in all the galaxies, 
and in particular the size \emph{difference} between blue and red GCs in 
NGC~4472 and NGC~4486 persists at all radii.  This same conclusion was 
reached for NGC~4472 by \citet{puz99}, although they did not estimate the 
absolute linear size of the GCs, and by \citet{kuea99} for the inner 
regions of NGC~4486. 

  In summary, the observable GC properties (colors, sizes) appear to be
nearly independent of position within the galaxies, at least in the three
large ellipticals studied here. Any changes in the average properties of
GCs result primarily from different mixtures of the two populations.
Unless the orbits of blue and red GCs are dramatically different, the
mechanism that was responsible for their size difference must have operated
with equal efficiency at all radii.

\subsection{Sizes of globular clusters in the Milky Way}

  As noted in Sect.~\ref{sec:sizes}, globular clusters in the Milky Way
show the same correlation between size and color as in other galaxies
(Table~\ref{tab:sizes}).  This is also illustrated in Fig.~\ref{fig:feh_rh_mw} 
where the effective radii are plotted as a function of metallicity and in
Fig.~\ref{fig:szdist_mw} which shows the size distributions for metal-rich 
and metal-poor Milky Way globulars (dividing at $\feh = -1$).  The two 
plots are very similar to those for extragalactic GCs in Figs.~\ref{fig:visz} 
and \ref{fig:szdist}.  

  As in the other galaxies in our sample, the correlation between GC size 
and metallicity is also preserved when subdividing the metal-poor and 
metal-rich GC populations in the Milky Way.  The median sizes of Milky Way 
globular clusters in the intervals $\feh < -1.5$, $-1.5 <\feh < -1$, 
$-1.0 <\feh < -0.5$ and $-0.5 <\feh $ are 3.28 pc, 3.00 pc, 2.48 pc and 
2.06 pc, respectively. However, since the metal-poor (large) GCs are 
preferentially found at large galactocentric distances, it is hard to say 
whether the size-metallicity correlation follows from the size-$R_g$ 
correlation, or vice versa. One way to approach this question would be to 
look at metal-poor and metal-rich clusters occupying the same volume of 
space, although this would still not exclude the possibility that the 
different populations might have different orbital characteristics. Another 
problem is one of pure statistics -- with a total of 147 known Milky Way 
GCs, sub-samples of GCs quickly become very small.

  Table~\ref{tab:sz_r_mw} lists the sizes of Milky Way GCs in three
radial bins: $0<R_g<2$ kpc, $2<R_g<5$ kpc and $5<R_g<10$ kpc. Although
the numbers of clusters in each bin are small, the table suggests
that the size difference between metal-rich and metal-poor GCs also exists
at all radii in our Galaxy. However, as in the ellipticals, 
further information about the orbits of individual clusters is necessary
to tell whether or not the size difference could be due to dynamical
processes.

\section{Discussion}

\subsection{Correlations with host galaxy parameters}

  A relation between GC mean metallicity and parent galaxy luminosity was first 
suggested by \citet{vdb75} and subsequently supported by spectroscopic work 
\citep{bh91}. With the discovery of multiple GC populations, an obvious 
question is whether the GC metallicity vs.\ host galaxy luminosity and 
other relations are present for both red and blue GCs, for only one
population, or if the observed mean relations might even result just from 
different mixtures of two populations with roughly constant metallicities.  
Previous attempts to address this question \citep[e.g.][]{for97,bur01} have 
generally been based on compilations of literature data, lacking homogeneity 
in the choice of filter systems, data reduction procedures etc.  Here we 
reinvestigate correlations between properties of GC subpopulations and 
their host galaxies, based, for the first time, on a large homogeneous 
dataset.

  In Fig.~\ref{fig:ccplots} the \vio\ colors of the two GC subpopulations
returned by the KMM test (Table~\ref{tab:kmm})
are shown as a function of the various host galaxy parameters listed in 
Table~\ref{tab:props}: absolute $B$ magnitude, central velocity dispersion, 
\vio, \jko, and \bko\ colors, and Mg2 index.  We have chosen to present 
the various relations using the GC \vio\ colors directly instead of 
transforming these to metallicities.  It should be kept in mind that some of 
the low-luminosity galaxies contain rather few GCs, so that peaks in their 
\vio\ color distributions may not be very well determined.  The open dots 
in Fig.~\ref{fig:ccplots} indicate data for the Fornax galaxies (NGC~1399, 
NGC~1404) which were transformed from \bi\ to \vi . 

  The dashed lines in each panel of Fig.~\ref{fig:ccplots} represent 
least-squares fits to the data points. Slopes and Spearman rank correlation
tests for the various fits are listed in Table~\ref{tab:ccorr}.
The colors of \emph{both} red and blue GCs are correlated with host galaxy 
$M_B$ and central velocity dispersion ($\log\sigma_0$) with a $>90$\%
probability, while trends with the Mg2 
index and host galaxy colors are only weak and, in the case of Mg2, mostly 
driven by the poor GC system of NGC~4733. The correlations with host
galaxy $M_B$ and $\log \sigma_0$ are all significant at the $2-3\sigma$ 
level, and remain relatively unchanged even if the two cluster-poorest 
galaxies, NGC~3384 and NGC~4733 are omitted from the fits. Although 
correlations for the blue and red GC populations are found at about the 
same significance level, the slope is generally slightly steeper for the 
red GC correlations.

  A correlation between the color of the \emph{red} GCs and host galaxy
luminosity was already noticed by \citet{for97}. These authors
did not find any convincing correlation with the color of the \emph{blue} 
peak but only a large scatter.  \citet{for97} used a fairly heterogeneous
dataset with metallicities derived from a mixture of spectroscopy, \vi , 
\bi\ and Washington photometry, which might explain why they failed to 
detect any correlation between the metallicity of the metal-poor GC 
populations and host galaxy properties.  Based on a larger (but still 
somewhat heterogeneous) sample, \citet{bur01} detected a correlation
between host galaxy $M_V$ and the metallicity of the blue peak.
The slope of their $\feh - M_V$ fit ($-0.06\pm0.01$) is formally quite 
similar to that of our $\vio - M_B$ relation, assuming a conversion factor 
of about 3 between \vio\ and \feh\ \citep{kis98}. However, excluding the 
cluster-poorest galaxies from their sample, \citet{bur01} found that the
slope of their $\feh$ vs.\ $M_V$ relation was reduced to $-0.02\pm0.02$.
They did not look at the red GCs.

  Based on a compilation of literature data, \citet{for00} detected a 
$3\sigma$ correlation between $\log \sigma_0$ and the color of the red peak. 
The slope of their relation (0.23) is quite similar to the one found here.
\citet{for00} found no significant correlation between $\log \sigma_0$ and 
the color of the blue peak, but their data comfortably allow a slope 
similar to the one indicated in our Fig.~\ref{fig:ccplots}. Finally, we
note that \citet{kun99} found the colors of both red and blue GCs to be
at best weakly correlated with host galaxy luminosity.

  One of the more remarkable features of Fig.~\ref{fig:ccplots} is the lack
of any clear correlation between host galaxy \emph{colors} and the GC colors. 
In 
\emph{most} galaxies, the colors of the red GCs roughly match the \vio\ color 
of the galaxy halo light, but galaxy colors span a wider range than the 
red GC populations and several galaxies have much redder integrated \vio\ 
colors than the red GCs. As an illustration of this, we have superimposed
a dashed-dotted line corresponding to a 1:1 correspondance between GC
and host galaxy \vi\ colors on Fig.~\ref{fig:ccplots}. Such a relation is 
clearly incompatible with the data.  At best, we see a weak correlation 
between the host galaxy \jko\ color and the color of the \emph{red} GC 
population, but a slope of 0 is within the $1.5\,\sigma$ errors even for 
this relation.  Since it is well-known that a relation exists 
between the integrated color and the luminosity of early-type galaxies 
\citep[e.g.][]{vdb75,sv78}, it is somewhat puzzling that we do not detect 
correlations between the GC colors and \emph{both} host galaxy luminosity 
and color.  In 
fact, a linear fit to the host galaxy \vio\ and $M_B$ values for the 
galaxies in our sample, listed in Table~\ref{tab:props}, yields 
$\vio = (-0.046\pm0.007) \, M_B + 0.253$.  If this is combined with the 
slopes of $-0.016\pm0.005$ and $-0.020\pm0.008$ for the GC color vs.\ host 
galaxy $M_B$ fits (Table~\ref{tab:ccorr}) then one would expect slopes of 
$0.35\pm0.11$ and $0.43\pm0.17$ for the GC color vs.\ host galaxy \vio\ 
relations. For the blue peak this is actually compatible with the measured 
slope within the $1\sigma$ errors, while there is a $\sim2\sigma$ discrepancy 
for the red peak.  The main reason for this is probably that the galaxy 
color--luminosity relations exhibit significant scatter. For example, 
inspection of Table~\ref{tab:props} shows that NGC~4472 has very 
blue \vio\ and \jko\ colors for its high luminosity, while the much less 
luminous NGC~4473 has the \emph{reddest} integrated \vio\ colors of all the
galaxies in the sample. Thus, globular cluster colors appear to be determined 
primarily by the host galaxy luminosity, rather than by whatever processes 
gave the field stars their colors. 


\subsection{GC subpopulations and formation scenarios}

  The presence of a correlation between GC colors and host galaxy
luminosities for both red and blue GCs would be a strong indicator that both 
GCs populations ``knew'' about the galaxy in which they formed. However, 
the lack of any obvious correlation between host galaxy \vi\ \emph{color} 
and the GC colors potentially provides us with an equally strong clue that 
the cluster- and star formation histories of galaxies may have been 
significantly different. This should come as no surprise -- as is well
known, the Milky Way is still an actively star forming galaxy, while
globular cluster formation ceased long ago in our Galaxy.

  The discovery of bimodal color distributions in many galaxies has led 
various authors to speculate about formation scenarios that could create 
distinct GC subpopulations. As mentioned in the introduction, the merger 
model by \citet{ash92} was the first to explicitly predict bimodal colors 
distributions. In this model, elliptical galaxies form by mergers of 
gas-rich spirals, creating the metal-rich GC population in the starburst 
associated with the merger. The metal-poor clusters are inherited from 
the progenitor galaxies. One major problem with this model, however, is 
the fact that spirals generally contain only few globular clusters, and 
that most large ellipticals are rich in both metal-rich and metal-poor 
GCs. Gas-rich mergers might account for relatively cluster-poor ellipticals,
but the exceedingly rich GC systems of galaxies like NGC~4486 and
NGC~4472 are very hard to explain within the merger picture \citep{harris99}.

  Alternatively, it has been suggested that the \emph{metal-rich} GCs
in ellipticals represent the galaxies' original GCs, while the metal-poor
ones have been accreted from smaller galaxies \citep{cot98} and/or formed
in minor mergers. Their model is able to successfully account for quite 
a wide variety of final metallicity distributions.  A somewhat similar 
approach was taken by \citet{hil99} who suggested that the metal-poor GCs 
in ellipticals formed by accretion of gas-rich dwarfs. One remaining problem 
with this approach is to verify that it is actually possible to 
accrete the required larger numbers of GCs without accreting any appreciable 
number of metal-poor \emph{field stars} at the same time.

  It is, however, still not clear how to explain the wide range in observed 
properties of GC color distributions. Some galaxies exhibit strikingly 
bimodal color distributions with roughly equal numbers of blue and red
GCs (like e.g.\ NGC~1404, NGC~4649, NGC~4472), while others show a much
reduced number of red GCs (NGC~4406) or even just a single (but still 
significantly broadened) metal-poor component (NGC~4365), and still others 
appear to show a fairly continuous color distribution, spanning a range 
similar to that observed in the truly bimodal systems (NGC~4552). 

If mergers or accretion played an important role, one might expect a 
dependence on environment.  In Fig.~\ref{fig:rho} we plot the two indicators 
of bimodality, $P$(kmm) and $P$(dip) as a function of galaxy density 
\citep{tul88}. For $P$(kmm), a low value (close to 0) indicates that a 
two-component fit is a major improvement relative to a one-component fit. 
For $P$(dip), a high value (close to 1) indicates a high probability that 
the dataset is not unimodal. Although current galaxy density is a quite
crude measure of past mergers, neither of the two plots shows any significant 
correlation with galaxy density, suggesting that the population mixture 
in GC systems is mostly shaped by intrinsic processes in the galaxies.

One possible signature of past mergers is kinematically distinct cores (KDCs).
Among the galaxies in our sample, several have KDCs. However, the GC systems
of these galaxies do not generally show strongly bimodal color distributions. 
In fact, one of these galaxies is NGC~4365 whose GC system appears to be 
composed almost entirely of a single metal-poor population. We also note 
that \citet{forb96} found no evidence for any connection between KDCs in 
galaxies and bimodality.

  Both mergers and accretion processes obviously take place even at the
current epoch and it is clear that the starbursts associated with these
events often lead to the formation of large numbers of very luminous
(and therefore presumably massive) star clusters.  However, the 
question is whether they were the dominating factors shaping the properties 
of GC systems in large ellipticals. It is now clear that massive star
clusters can form under a wide variety of circumstances \citep{lr00} and 
the high levels of star formation in the gas-rich environments in
proto-galaxies may quite naturally have led to formation of globular clusters.
The observation that the colors of \emph{both} GC populations appear to 
correlate with host galaxy properties would be hard to explain within the 
accretion / merger pictures and would fit better into \emph{in situ}-type 
scenarios in which all GCs ``knew'' about the size of the final galaxy to 
which they would eventually belong.  In order for this to be possible, the 
initial episodes of GC formation in giant ellipticals (gEs) must have 
taken place \emph{after} they assembled into individual entities, although not 
necessarily having evolved into anything we would recognize as a gE today. 

  Such a scenario was outlined by \citet{for97}, who further proposed that
the initial episode of GC formation would be halted by the dispersion of
gas by supernova explosions.  A few Gyrs later, as the gas cools down and 
falls deeper into the potential well, star formation commences again, 
forming the second (metal-rich) generation of stars and GCs.  A very 
similar scenario was favored by \citet{hhp99} and \citet{hh00}, based on 
deep HST observations of the red giant branch in the nearby giant elliptical 
NGC~5128. They found an extremely broad metallicity distribution function
for field stars in this galaxy, extending from $\feh \sim -2$ to at least 
solar, which was remarkably well matched by a two-component model of 
simple closed-box chemical evolution.  

  One might argue that the distinction between the various formation scenarios 
is somewhat blurred at the earliest epochs. \citet{hhp99} point out that
the first generation of star formation within the halo of a proto-gE may well 
have proceeded within a large number of relatively distinct gaseous clumps.  
In a proto-galactic halo with 
numerous gaseous fragments, many of these fragments are bound to interact 
and form larger subunits. Such events would share many of the characteristics 
of ``mergers'', especially when two relatively large subunits collide. In 
some occasions, such large-scale events could lead to a major burst of GC 
formation in which much of the remaining gas would be consumed, forming a 
quite distinct second peak in the metallicity distribution as observed 
in e.g.\ NGC~4649. 

  Even if some characteristics of GC systems might be understood within
a framework like the one outlined above, many observed properties evidently
remain to be accounted for. One remaining issue is to explain the 
\emph{size} difference 
between GC subpopulations.  This phenomenon seems 
to be a quite ubiquitous one, observed in all galaxies with broad GC color 
distributions (even in the Milky Way!) and at all galactocentric distances.  
The size differences may be either primordial, or a result of dynamical 
evolution. The fact that the differences exist in all galaxies and at all 
radii would appear to indicate a high probability that 
the differences are actually set up at formation.  Since more compact 
clusters presumably originated from denser proto-globular gas clouds, this 
may hold information about the physical condition of the gas phase and 
star formation processes at the time of GC formation.  The physics 
of cluster formation is still very poorly understood, but observations of 
young globular-like clusters in nearby galaxies may very well hold important 
clues to a better understanding of this important issue. Another challenge
is to understand the lack of any significant correlation between
host galaxy colors and GC colors. One could speculate that globular clusters
form preferentially during the first, burst-like phases of major star forming
episodes, while field stars continue to form in residual, enriched gas
which was not used up initially in much the same way as stars are forming in 
the Milky Way today.  However, a better understanding of
star forming processes, particularly in the early Universe, is necessary 
before such ideas can be put on firmer ground.

\section{Summary and conclusions}

  Using deep HST / WFPC2 data, we have performed a detailed analysis 
of the color, size and luminosity distributions of globular clusters in 
17 nearby galaxies.  The main results may be summarized as follows:

\begin{itemize}
  \item In all but a few cases, a KMM test finds that two Gaussians 
        provide a significantly improved fit to the color distribution
        relative to a single Gaussian. Simple histograms of the color
        distributions or a DIP test do not always confirm significant 
        bimodality even in cases where the KMM test returns a very
        high confidence level for a two-component fit, but the peaks
        in the \vio\ color distributions found by KMM are nevertheless
        quite consistent even in the less obviously bimodal cases.
        
  \item Red GCs are generally smaller than blue GCs by typically $\sim 20$\% .
        This is true also in galaxies without sharply defined bimodal
        color distributions and within ``sub-bins'' in color.  In all 
        galaxies, GC sizes decrease as a function of \vio\ color in the 
        range $0.70 < \vio < 1.20$, while the sizes of the very reddest 
        clusters ($\vio > 1.20$) are again larger in some galaxies.  When 
        subdividing the clusters into only two broad bins, this effect 
        can sometimes conceal a size difference between red and blue GCs.
        
        The GC size difference exists not only in ellipticals and S0
        galaxies, but also in the one Sa galaxy (M104) in our sample.
        Perhaps even more remarkably, GCs in the \emph{Milky Way} follow the
        exact same trend.
               
  \item Fitting $t_5$ functions to the globular cluster luminosity functions,
        we find that the turn-over of the blue GCs is generally brighter
        than that of the red ones by about 0.4 mag on the average for the
	full sample.  This difference
        is slightly larger than that expected from population synthesis 
        models if the two populations have identical mass functions, ages 
        and stellar IMFs, but different metallicities. However, if the
	sample is restricted to galaxies for which the error on the
	turn-over difference is less than 0.25 mag, then the average 
	difference between the turn-overs of blue and red GCs is only
	0.3 mag, which is very close to the 0.26 mag expected for 
	similar ages and mass functions but different metallicities.
	The Milky Way and Sombrero spiral galaxies both show a similar
	offset of $\sim 0.5$ mag, but curiously, in M31 the \emph{blue} 
	(metal-poor) GCs seem to be fainter.

        Using luminosity instead of magnitude units, the upper part of the 
        GCLF ($L \ga 10^5 \lsun$) is generally well fit by a power-law with 
        exponent $\alpha \sim -1.75$. Thus, the mass function of old GCs 
	brighter than the turn-over is apparently very similar to that 
	of young cluster systems, and it seems plausible that old GCs
	and young clusters form by the same basic mechanism.
        
  \item We have apparently detected a second case of ``faint extended'' 
        clusters in the nearby S0-type galaxy NGC~3384, similar to those 
        in NGC~1023 \citep{lar00}. In the four galaxies in our sample
        where such objects are definitively detectable, they thus appear 
        to exist in two.  These extended clusters are generally fainter 
        than the GCLF turn-over and thus tend to raise the lower end of 
        the GCLF, making it deviate from the Milky Way GCLF.  However, most 
        of the galaxies in our sample are too distant to tell from current 
        data if they possess similar clusters, so it remains unknown how 
        common such objects are.
        
  \item In three cluster-rich galaxies with several pointings
        (NGC~4472, NGC~4486, NGC~4365) we have examined radial trends
        in GC size and color. The size difference in NGC~4472 and NGC~4486 
        between blue and red GCs exists at all radii and we find no 
        evidence for any correlation between either the \vio\ peak colors
	or GC sizes and distance from the galaxy centers in these two galaxies. 
	NGC~4486 shows a strong increase in the relative numbers of blue 
	and red clusters outwards. Within the HST fields, such a trend is 
	much weaker in NGC~4472.  NGC~4365 seems to host only one 
	(metal-poor) GC population with no strong color gradients, but GCs 
	in the outermost bins do tend to be larger in this galaxy.  

  \item We have investigated correlations between the colors of GC
        subpopulations and host galaxy properties. Both the blue and red peak
        \vio\ colors correlate at the $2-3\sigma$ level with host galaxy 
        luminosity and central velocity dispersion. However, there is no 
        evident correlation between GC colors and the colors of their host 
        galaxies. Likewise, indicators of bimodality (KMM and DIP tests) 
	show no correlation with surrounding galaxy density.  
\end{itemize}

  We conclude that our data are best explained within \emph{in-situ} 
formation scenarios in which both GC populations formed within the 
potential well of the proto-galaxy, possibly in multiple episodes of 
star formation.

\acknowledgments
  This work was supported by HST grants GO.05920.01-94A and GO.06554.01-95A,
  National Science Foundation grant number AST9900732, NATO grant CRG 971552 
  and Faculty Research funds from the University of California, Santa Cruz.
  We thank Markus Kissler-Patig for useful discussions and Ken Freeman
  for his help, and the helpful comments of an anonymous referee are
  appreciated.

\newpage

\epsfxsize=7cm
\epsfbox{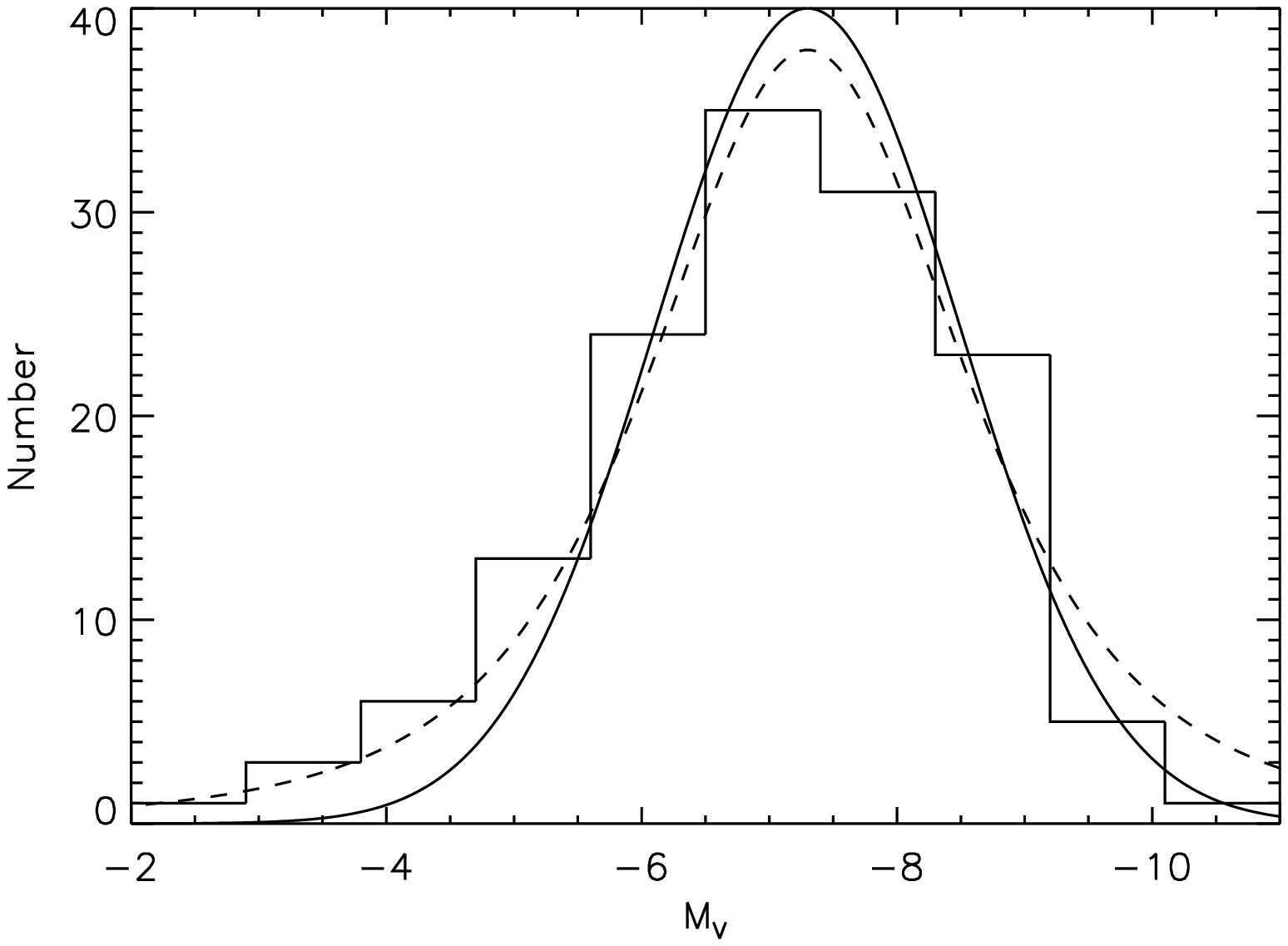}
\epsfxsize=7cm
\epsfbox{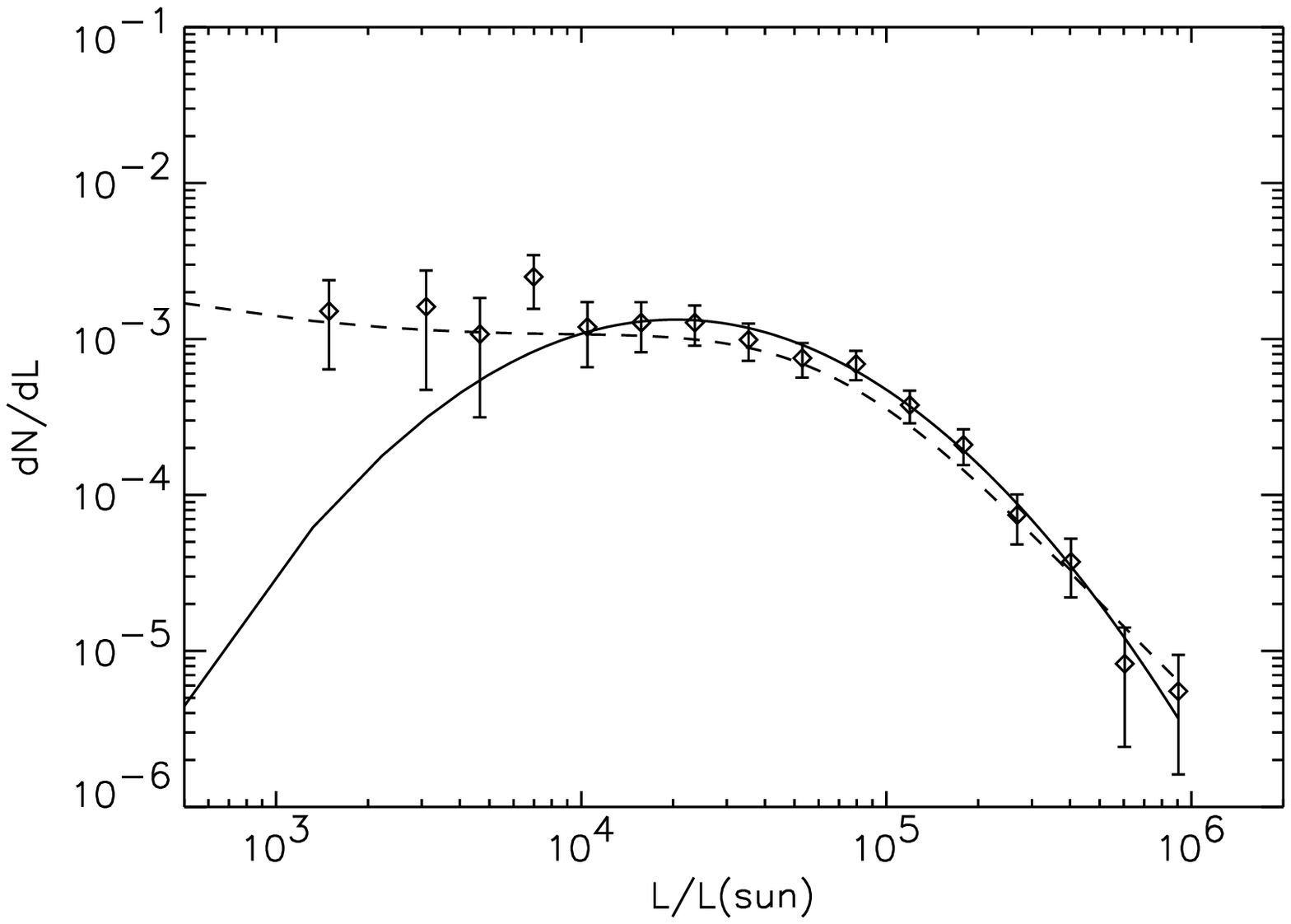}
\figcaption[Larsen.fig1a.ps,Larsen.fig1b.ps]{\label{fig:gclf_mw}
  Luminosity function in magnitude units (left) and luminosity units (right)
  for globular clusters in the Milky Way. Overplotted are Gaussian 
  (solid line) and $t_5$ function (dashed line) fits.
}

\noindent \epsfysize=19cm \epsfbox{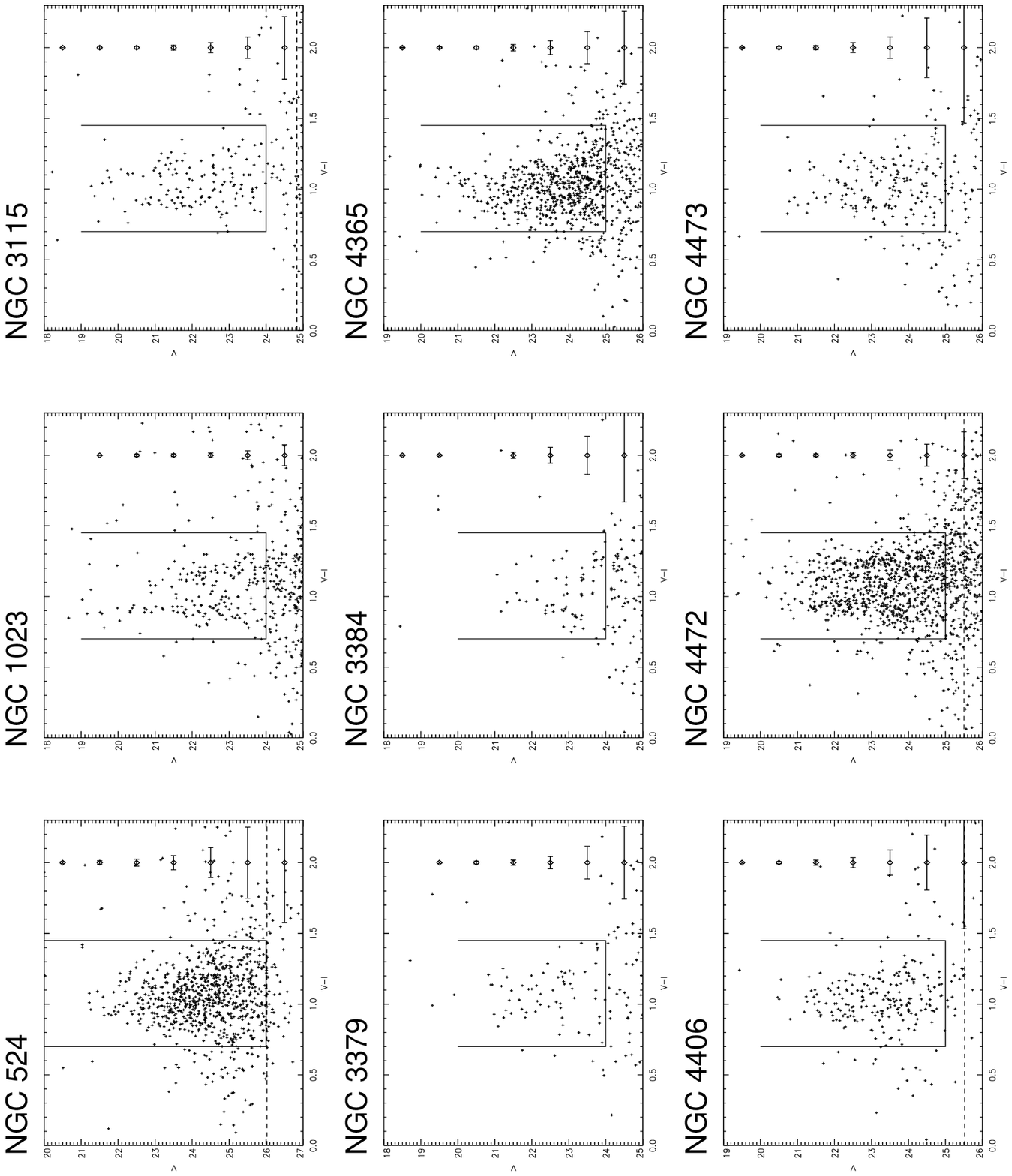}

\noindent \epsfysize=19cm \epsfbox{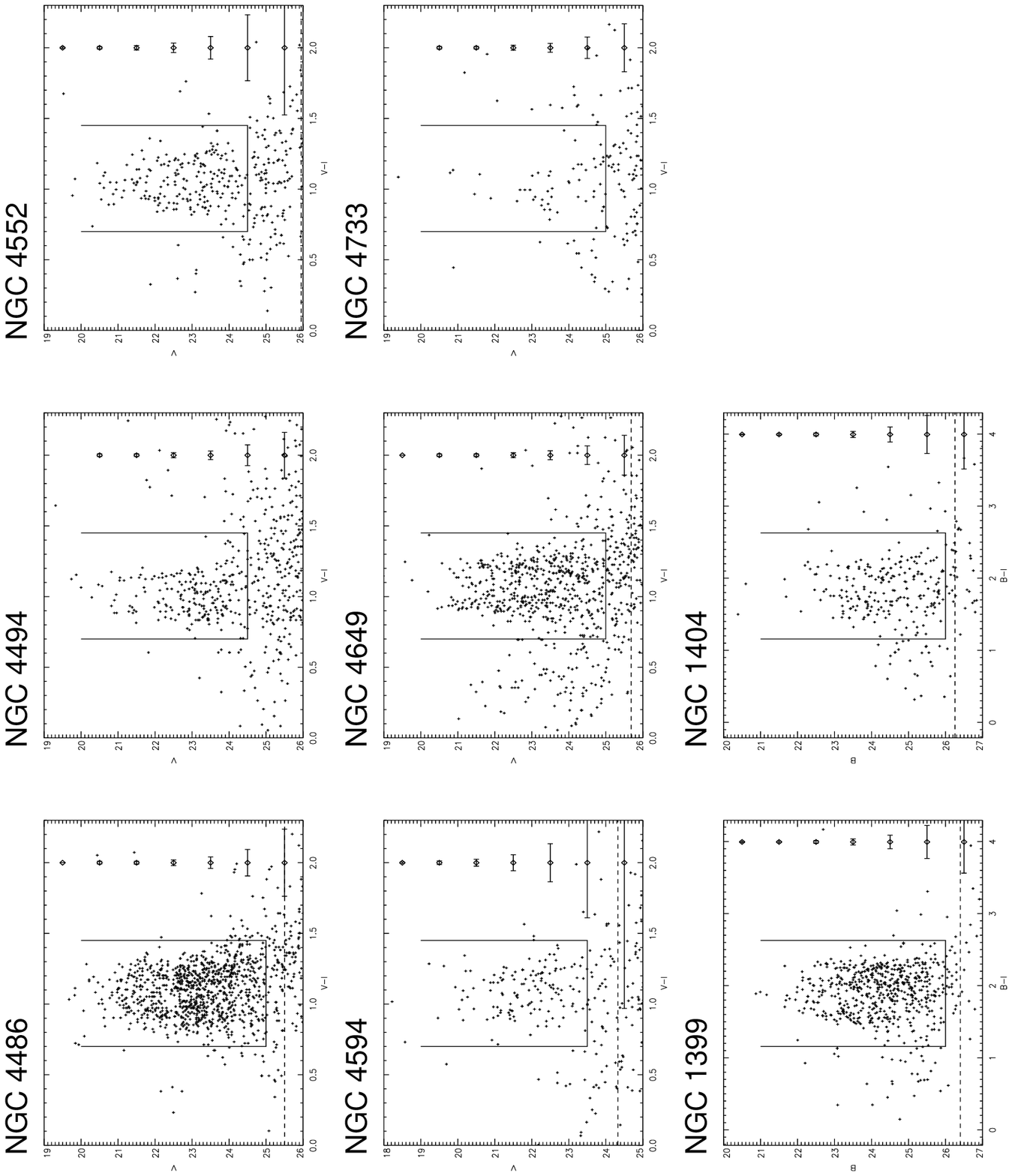}
\figcaption[Larsen.fig2a.ps,Larsen.fig2b.ps]{\label{fig:cmd}
  $\vio ,V$ color-magnitude diagrams for the galaxies. The boxes indicate
the part of the CMDs within which globular cluster candidates were selected
and the horizontal dashed lines indicate approximate 50\% completeness
limits.
}

\epsfxsize=7cm
\epsfbox{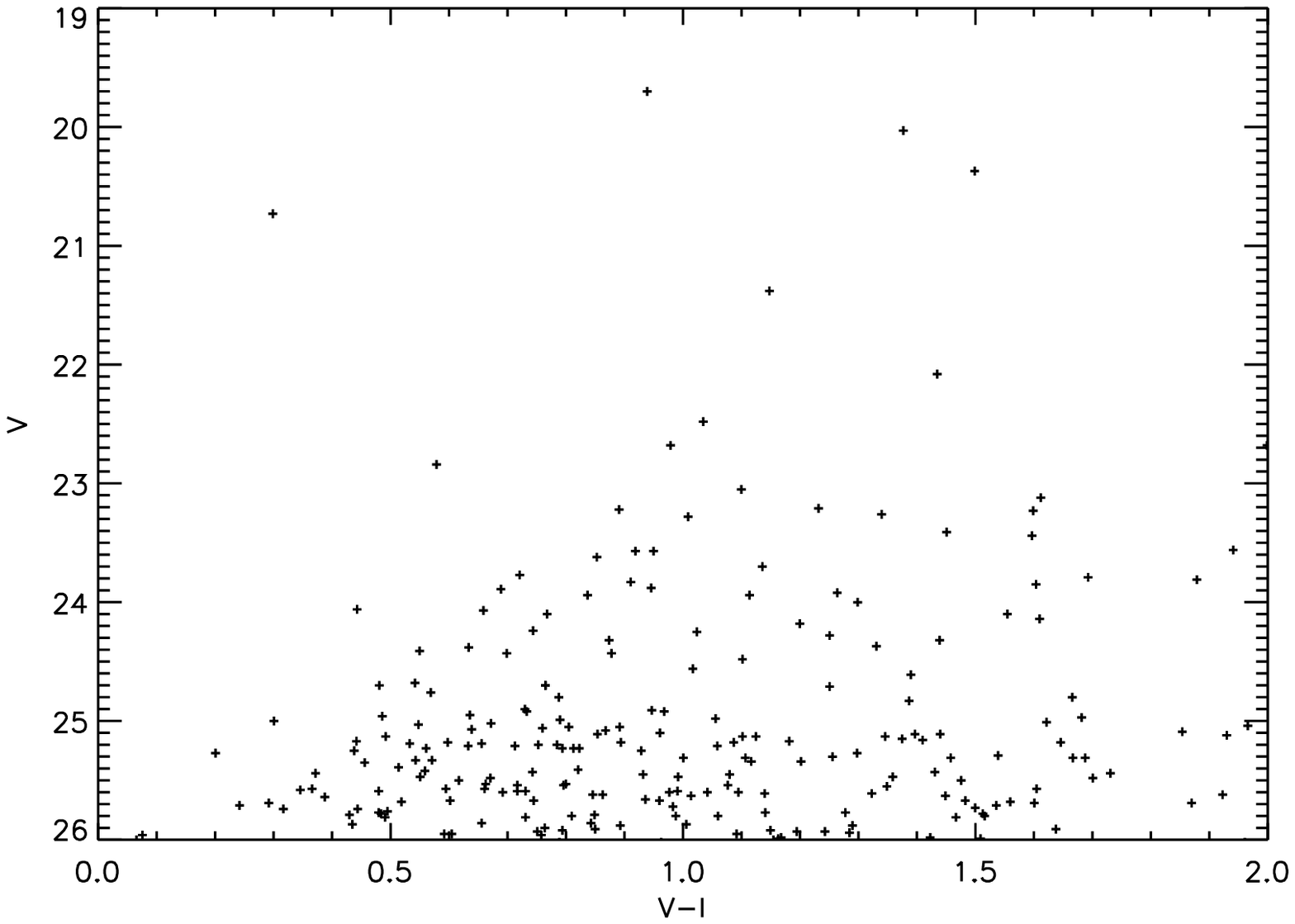}
\epsfxsize=7cm
\epsfbox{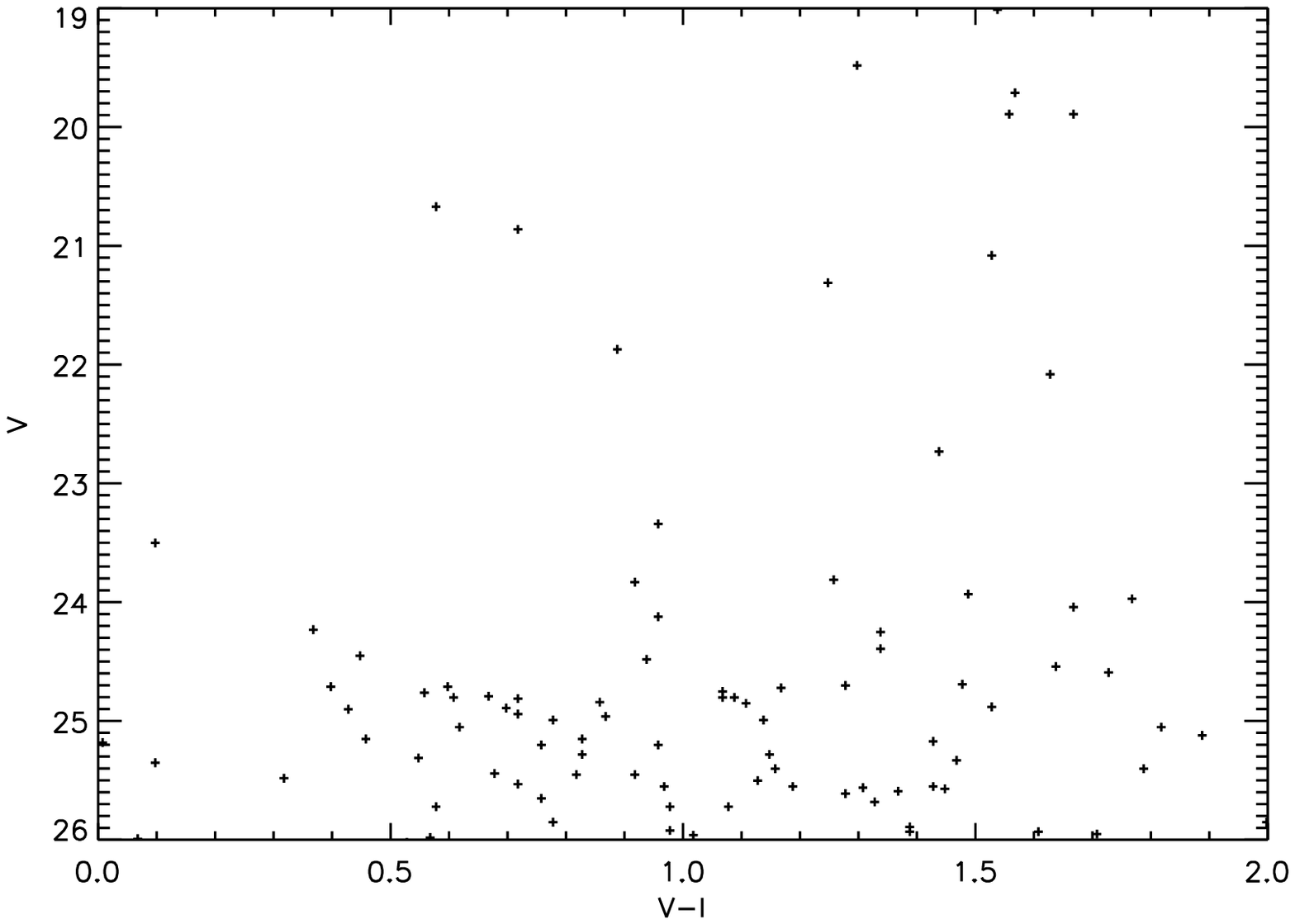}
\figcaption[Larsen.fig3a.ps,Larsen.fig3b.ps]{\label{fig:cmd_cmp}
  $(\vi)_0 ,V$ color-magnitude diagrams for the two comparison fields.
  Left: the Hubble Deep Field (North). Right: A field located about $2\deg$
  from NGC~1023.
}

\noindent \epsfbox{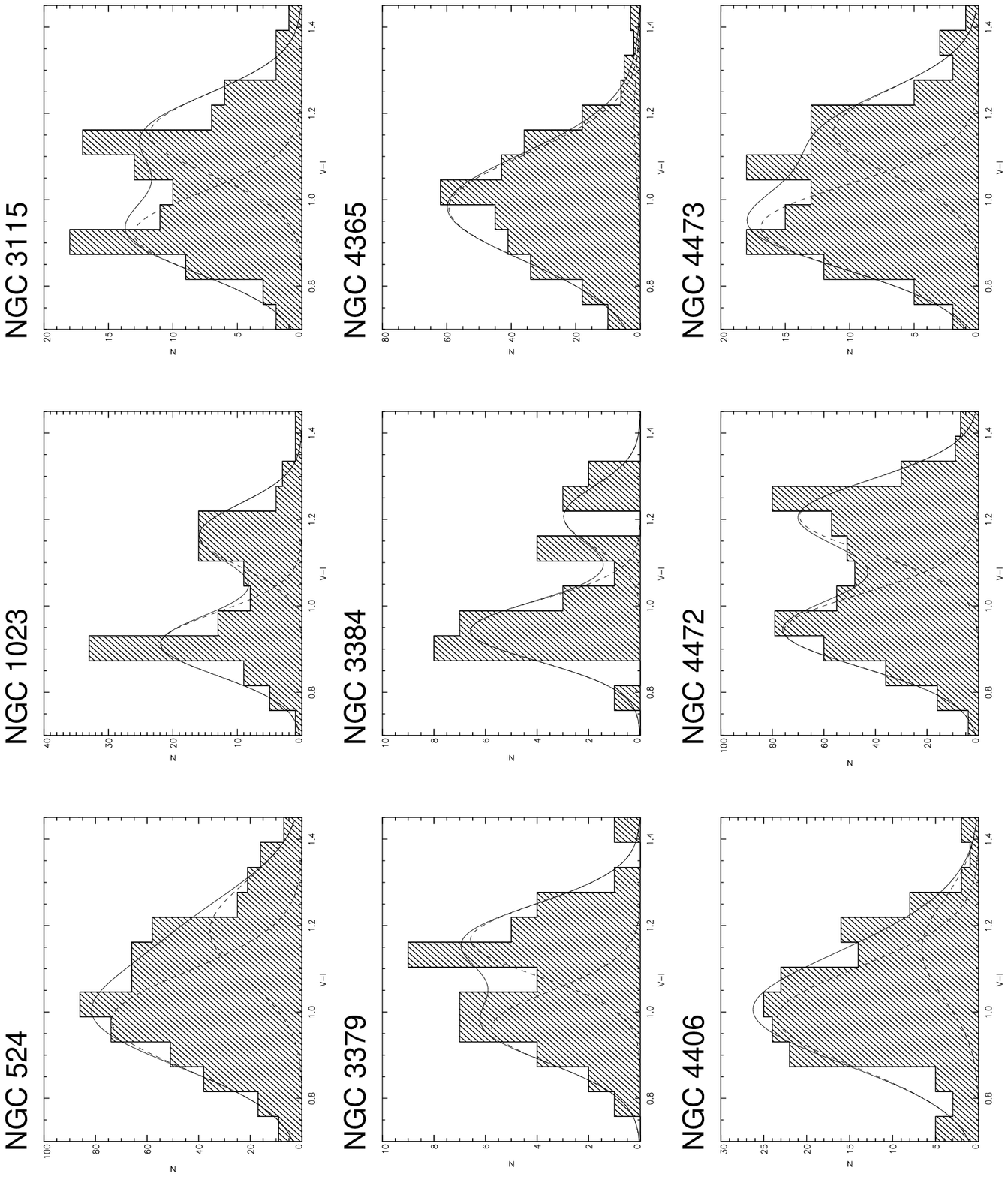}

\noindent \epsfbox{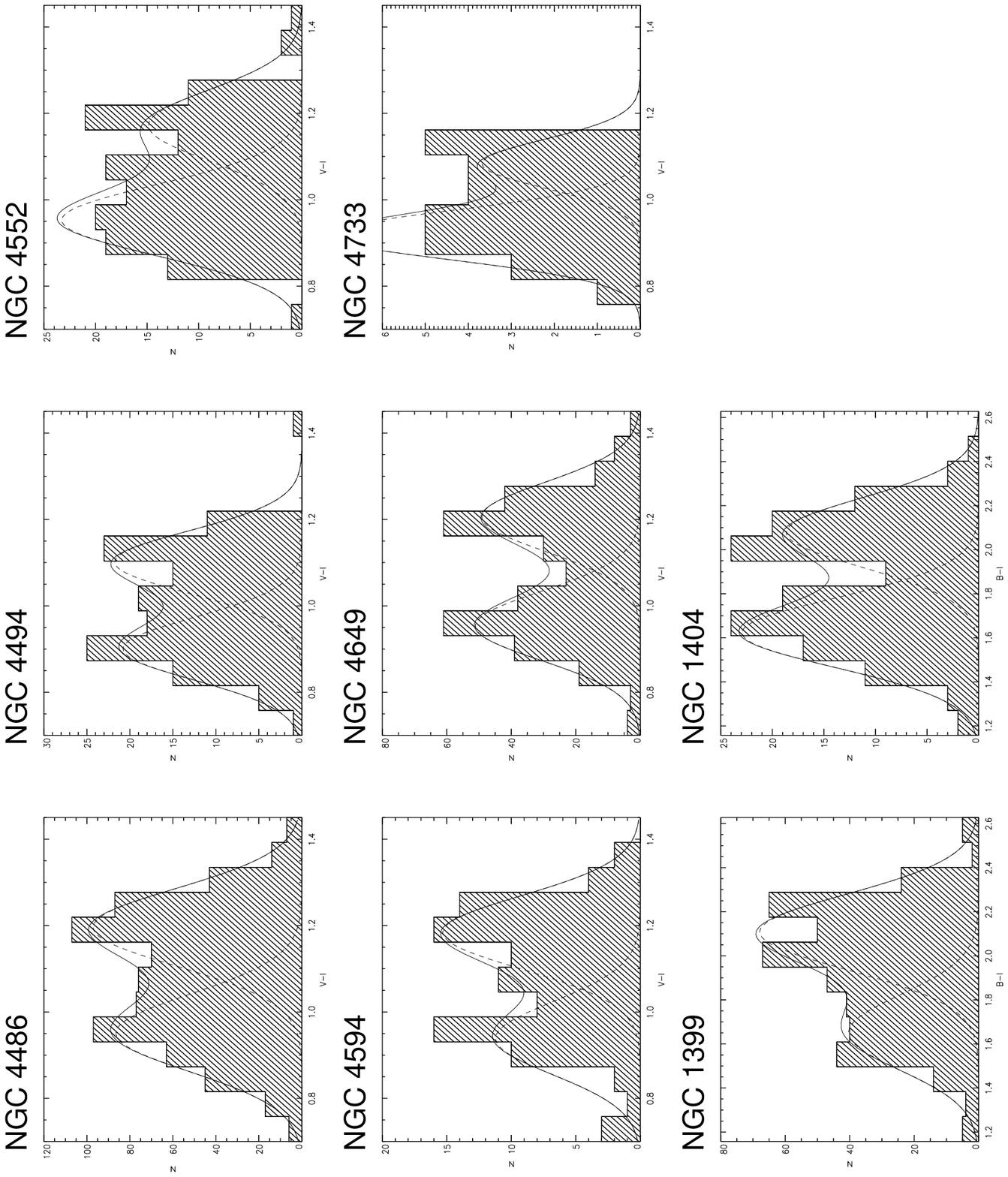}
\figcaption[Larsen.fig4a.ps,Larsen.fig4b.ps]{\label{fig:kmm}
  Histograms for the \vio\ color distributions of GCs.  Overplotted on 
the histograms are Gaussians corresponding to the two color peaks found 
by the KMM test and their sum.
}

\noindent \epsfbox{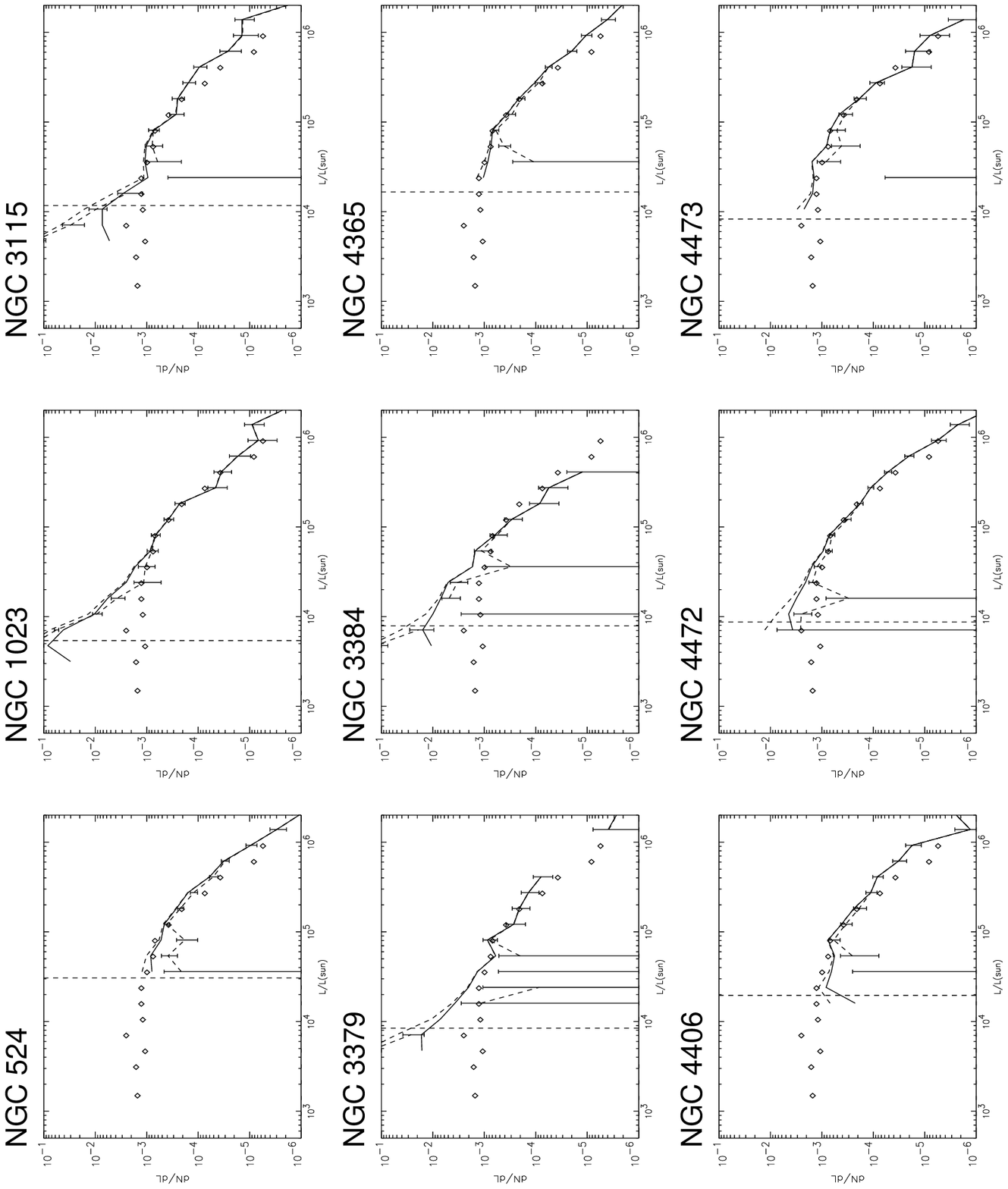}

\noindent \epsfysize=185mm \epsfbox{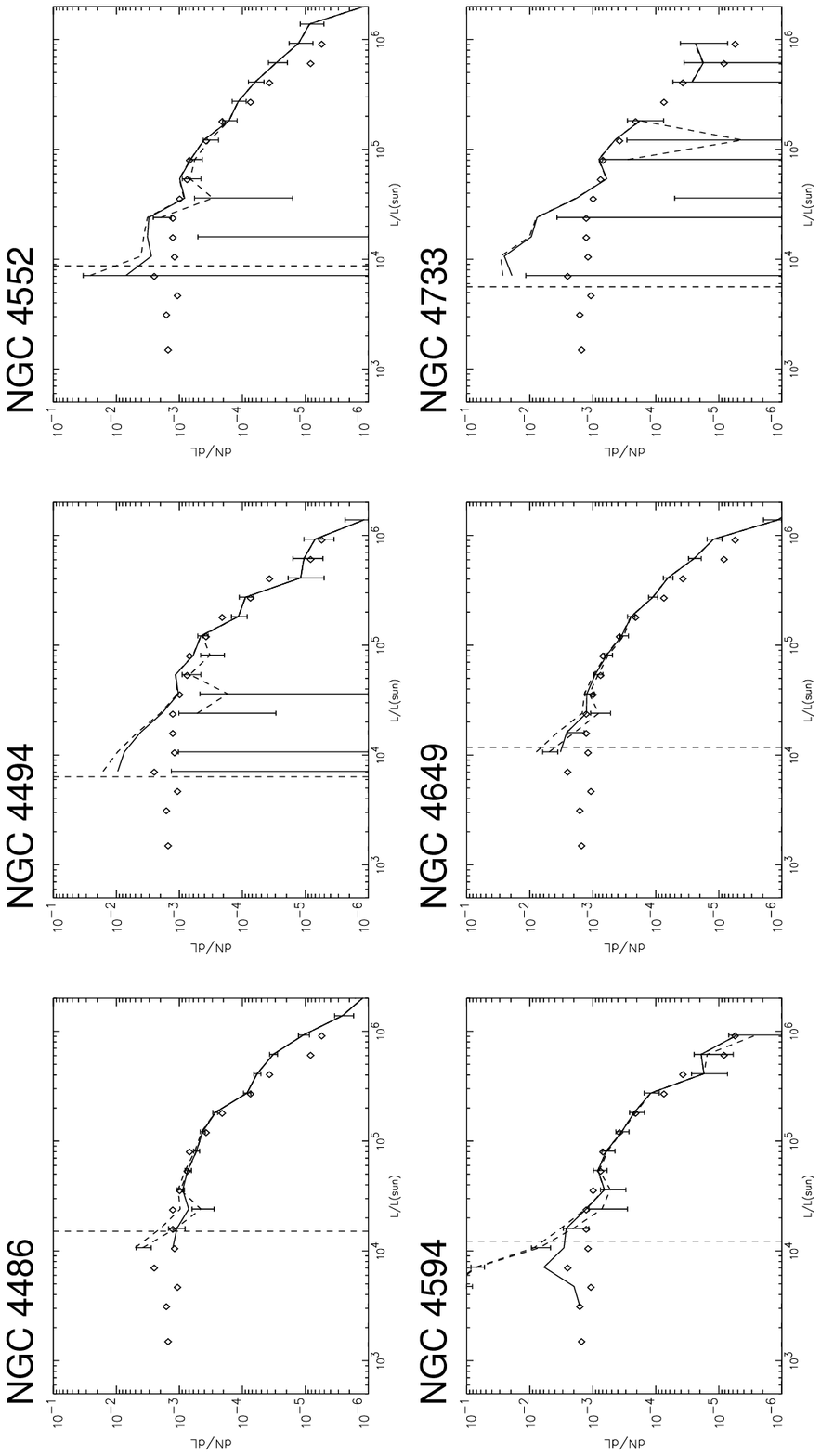}
\figcaption[Larsen.fig5a.ps,Larsen.fig5b.ps]{\label{fig:ldf}
  Luminosity functions for globular clusters in the 15 galaxies with
  $V,I$ photometry, compared with the Milky Way GCLF (dots). The
  various lines in each plot represent the same data, but different
  assumptions about completeness / contamination corrections (see text
  for details).  The turn-over 
  magnitude of $M_V = -7.5$ corresponds to about $8\times10^4 \lsun$.
}

\noindent \epsfbox{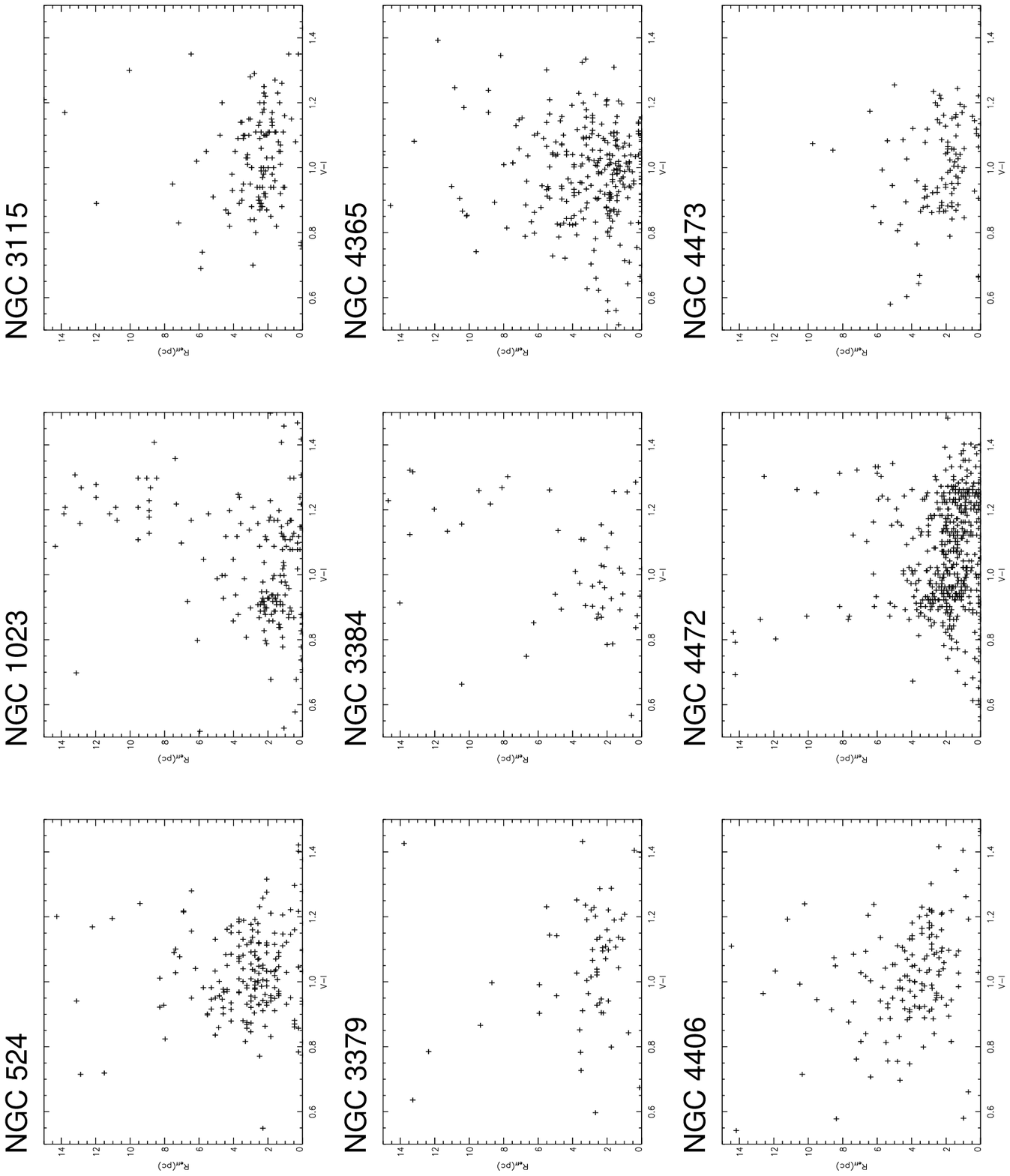}

\noindent \epsfbox{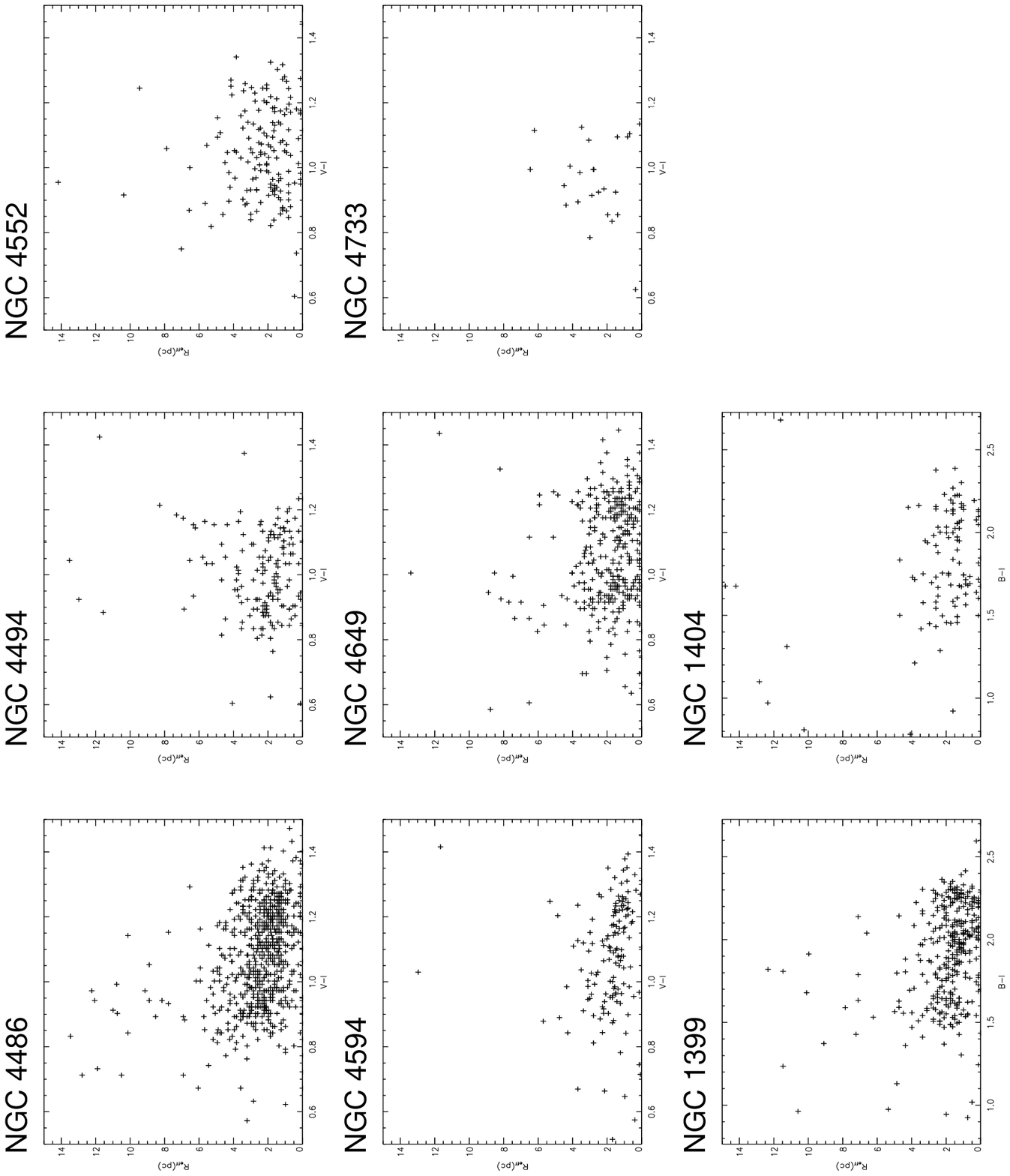}
\figcaption[Larsen.fig6a.ps,Larsen.fig6b.ps]{\label{fig:visz}
  Size as a function of color for globular clusters in the galaxies.
}

\noindent \epsfbox{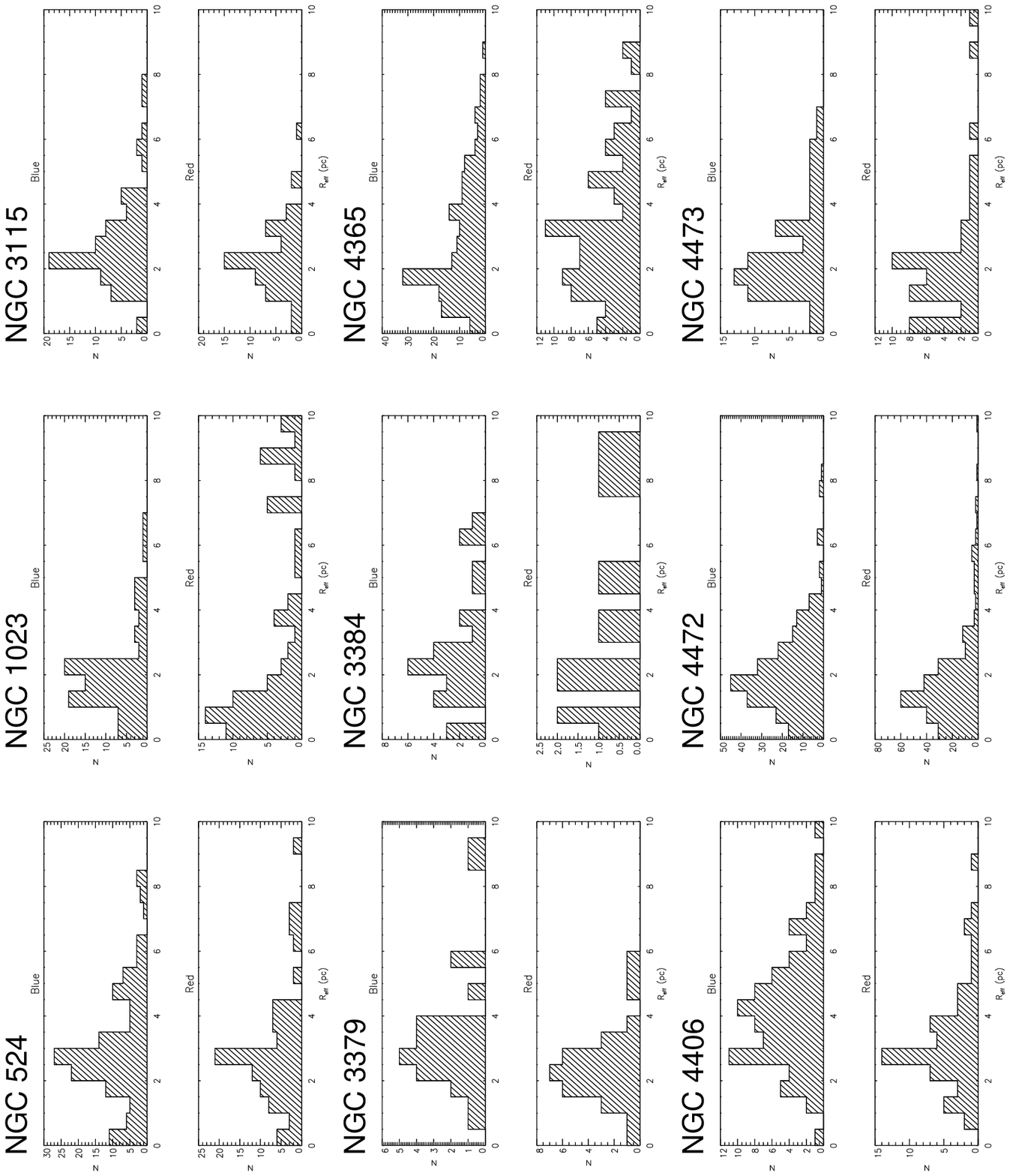}

\noindent \epsfbox{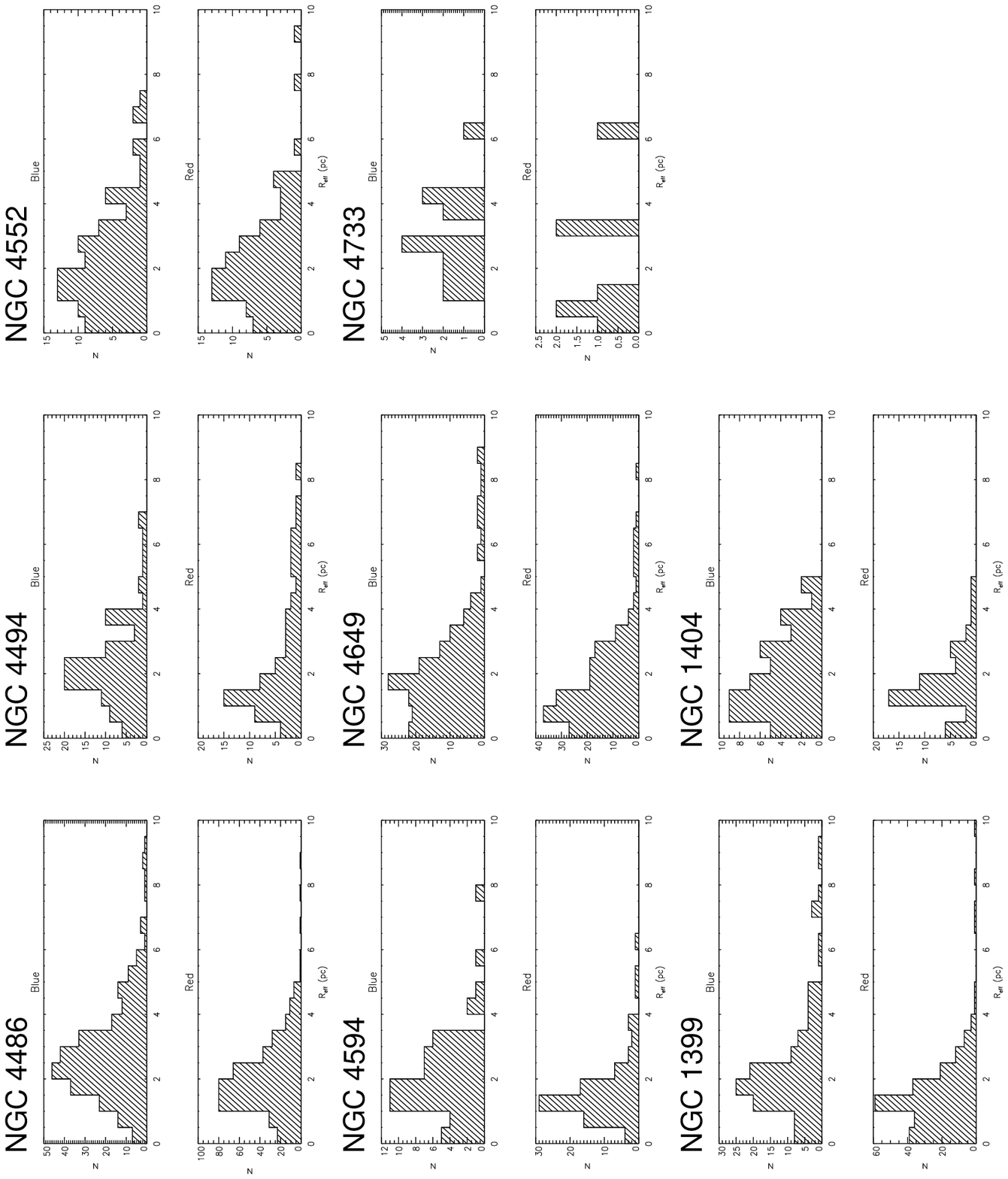}
\figcaption[Larsen.fig7a.ps,Larsen.fig7b.ps]{\label{fig:szdist}
  Size distributions for blue ($\vio <1.05$) and red ($\vio >1.05$) globular
  clusters. In each panel, the upper plot is for blue GCs and the lower
  plot is for red GCs.
}

\epsfxsize=10cm
\epsfbox{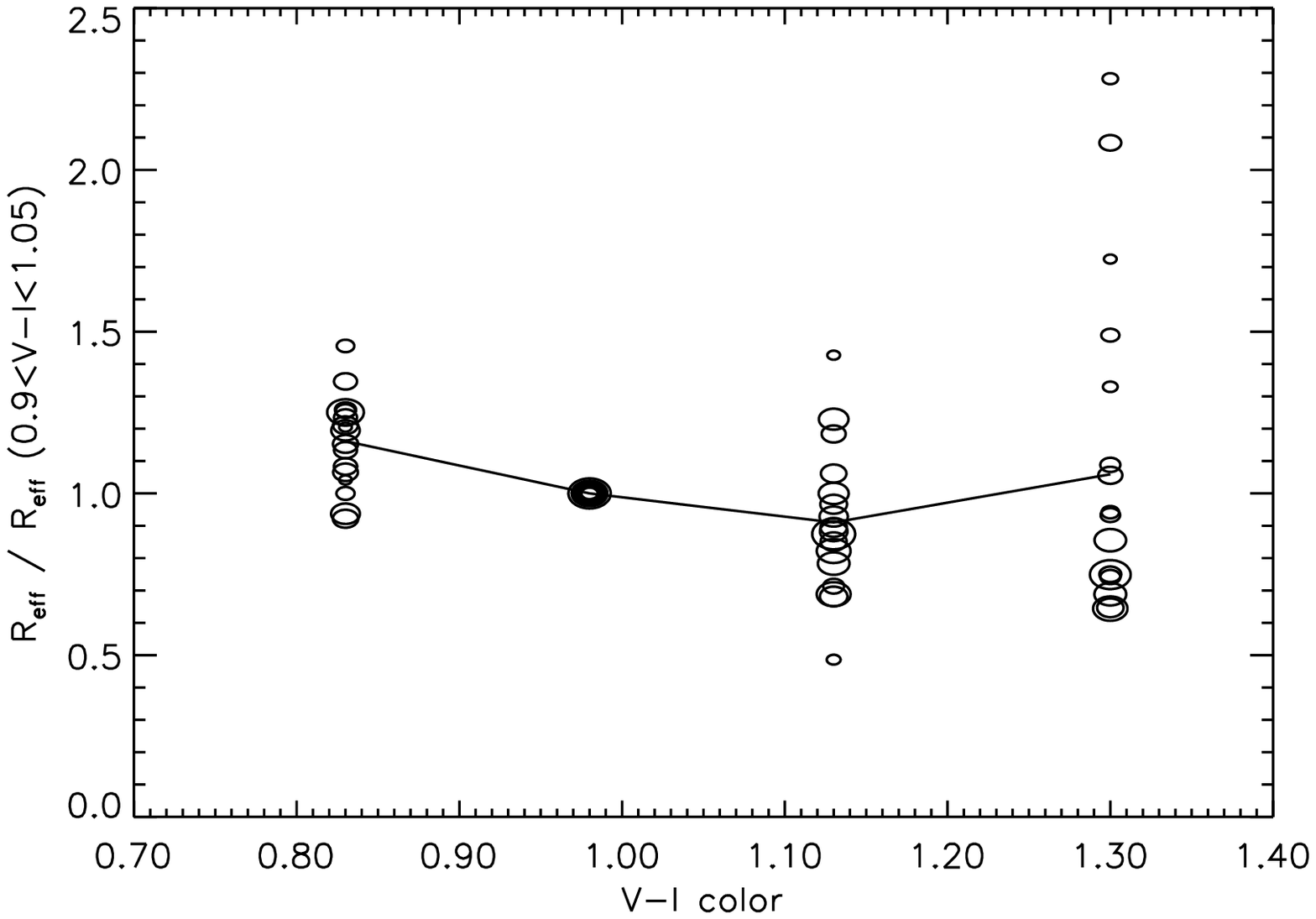}
\figcaption[Larsen.fig8.ps]{\label{fig:szplot}
  Comparison of GC sizes in four $(\vi)_0$ bins. Sizes have been normalized
to 1.0 in the $0.90<\vio<1.05$ bin. The symbol sizes are proportional to 
the logarithm of the number of GCs corresponding to each 
datapoint.  The line indicates the (unweighted) average of all data points 
at each bin.
}

\begin{minipage}{16cm}
\epsfxsize=15cm
\epsfbox{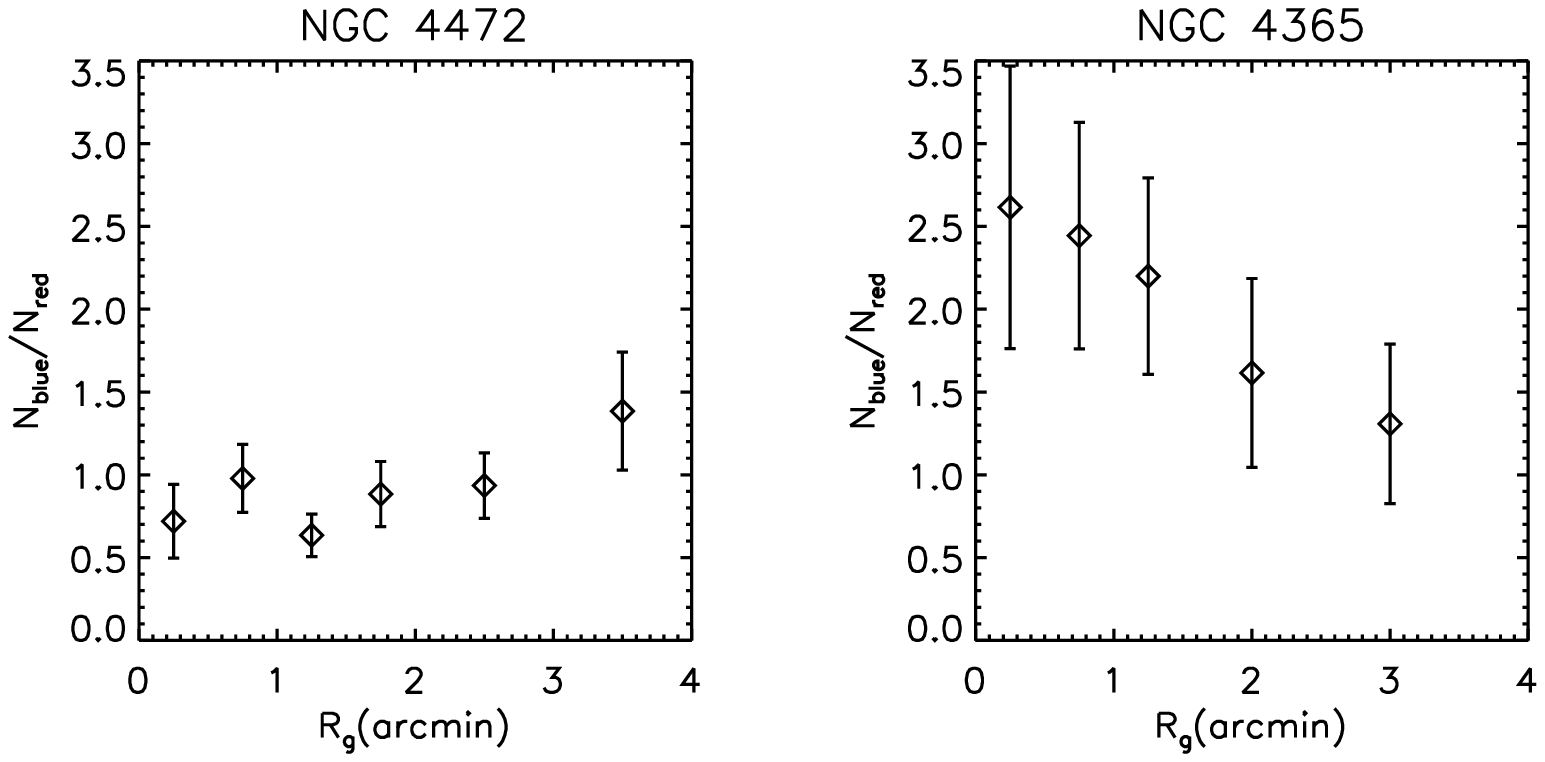}

\epsfxsize=15cm
\epsfbox{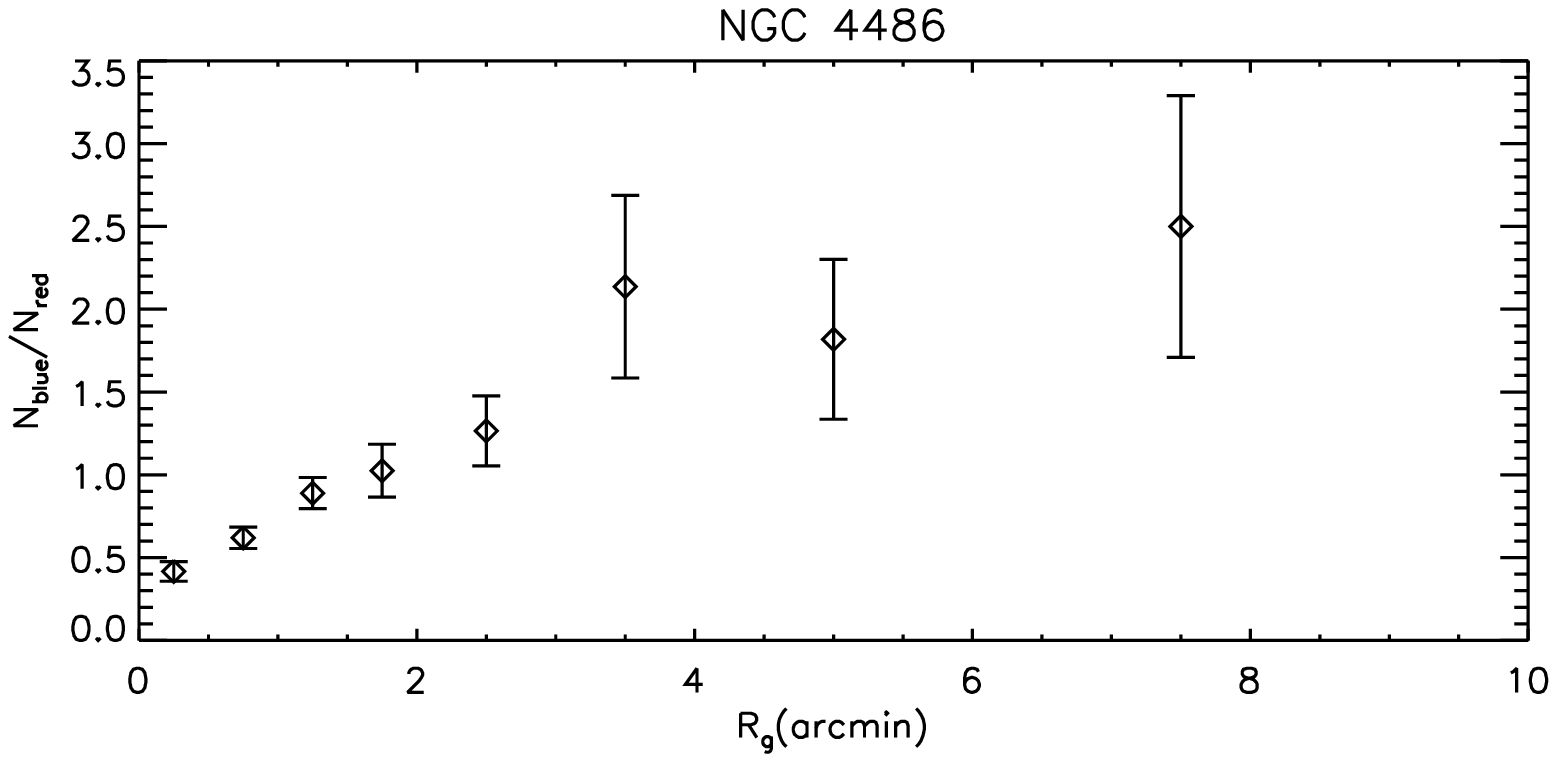}
\end{minipage}
\figcaption[Larsen.fig9a.ps,Larsen.fig9b.ps]{\label{fig:rplot}
  Ratio of the numbers of blue and red clusters as a function of 
galactocentric distance in arcmin ($R_g$) for NGC~4365, NGC~4472 (M49) and 
NGC~4486 (M87). At the distance of these galaxies, one arcmin equals 
about 4.6 kpc.
}

\epsfxsize=10cm
\epsfbox{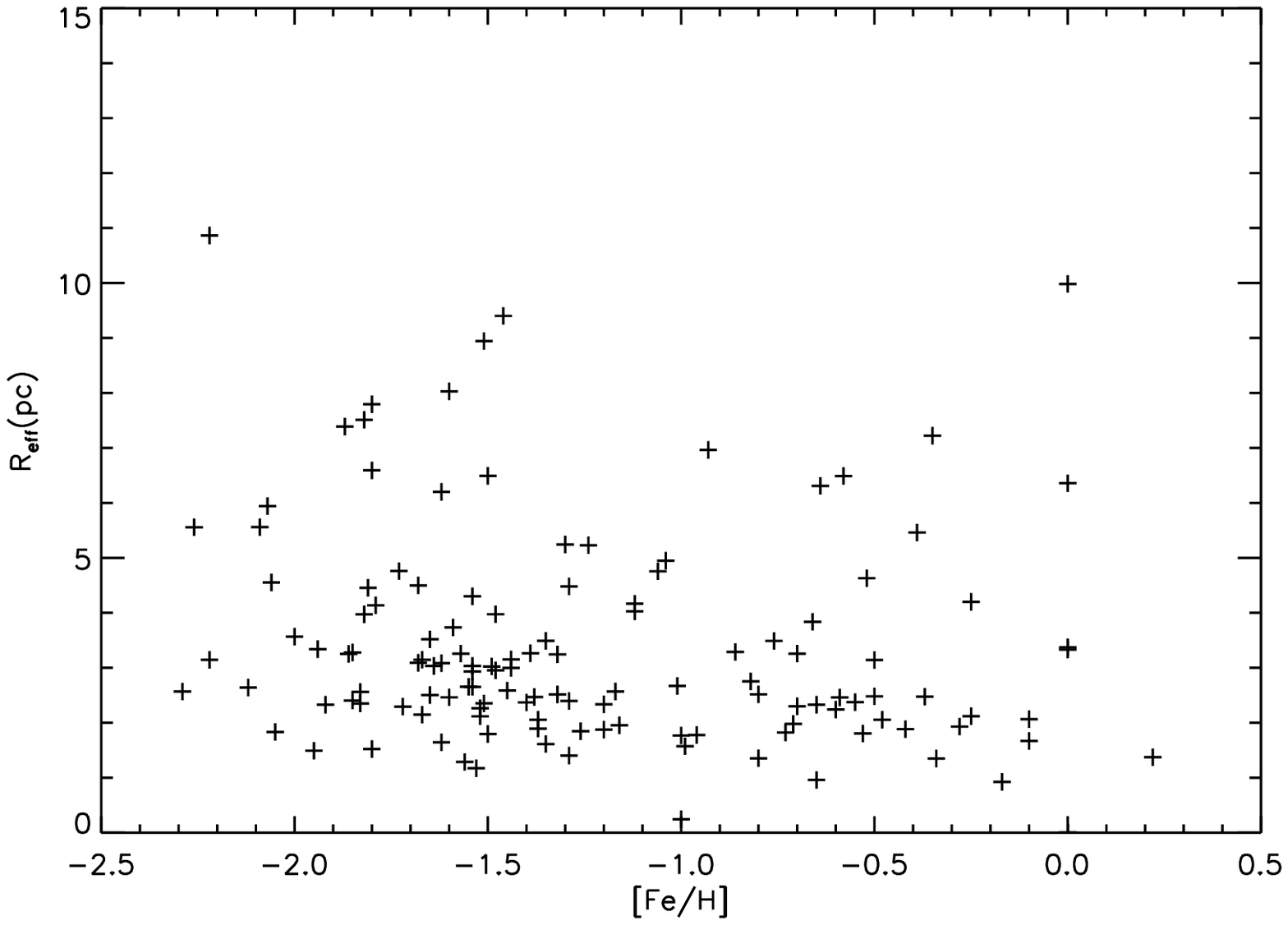}
\figcaption[Larsen.fig10.ps]{\label{fig:feh_rh_mw}
  Effective radii for Milky Way GCs as a function of metallicity.
}

\epsfxsize=10cm
\epsfbox{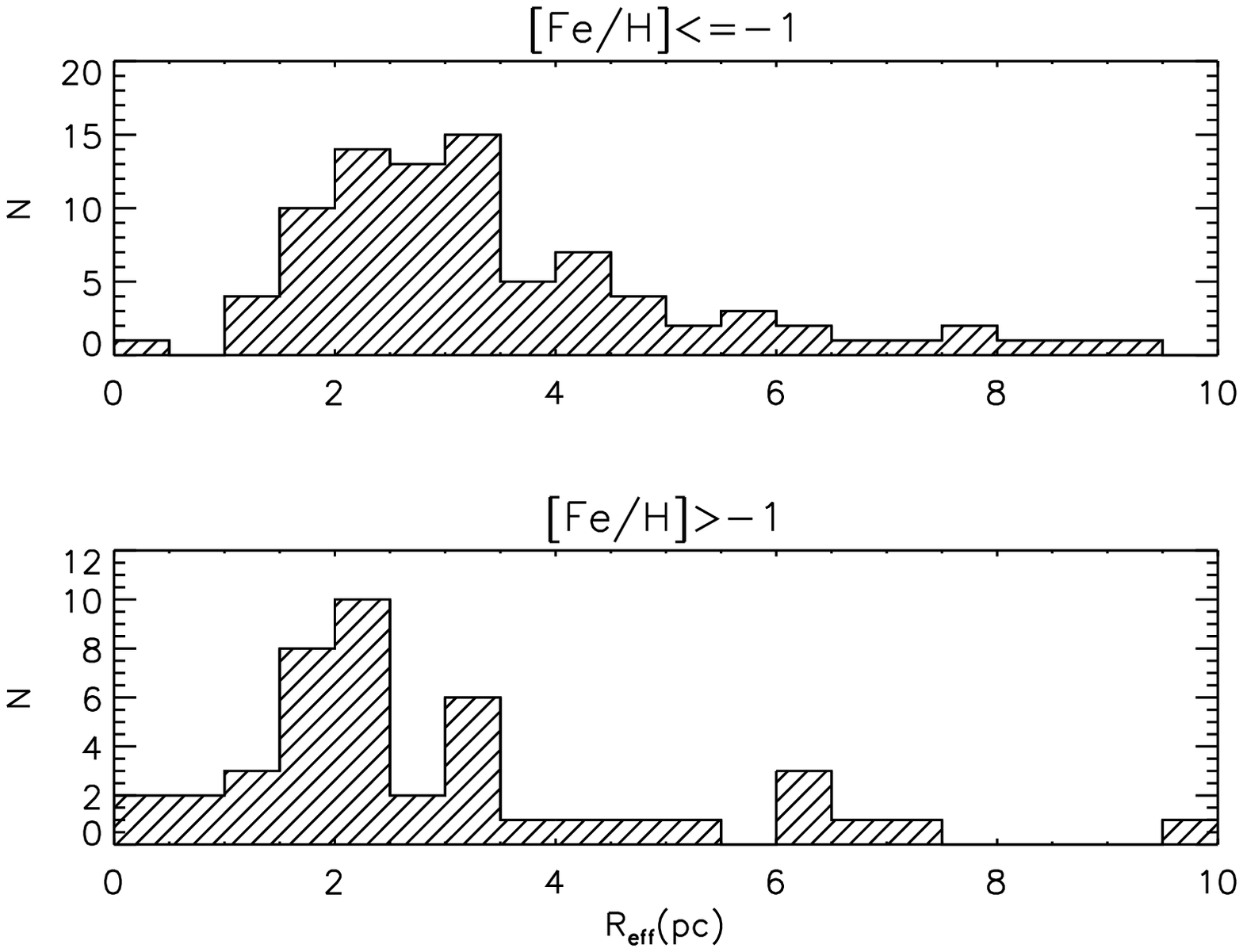}
\figcaption[Larsen.fig11.ps]{\label{fig:szdist_mw}
  Size distribution for metal-poor ($\feh <-1$) and metal-rich ($\feh >-1$) 
  globular clusters in the Milky Way.
}

\epsfxsize=16cm
\epsfbox{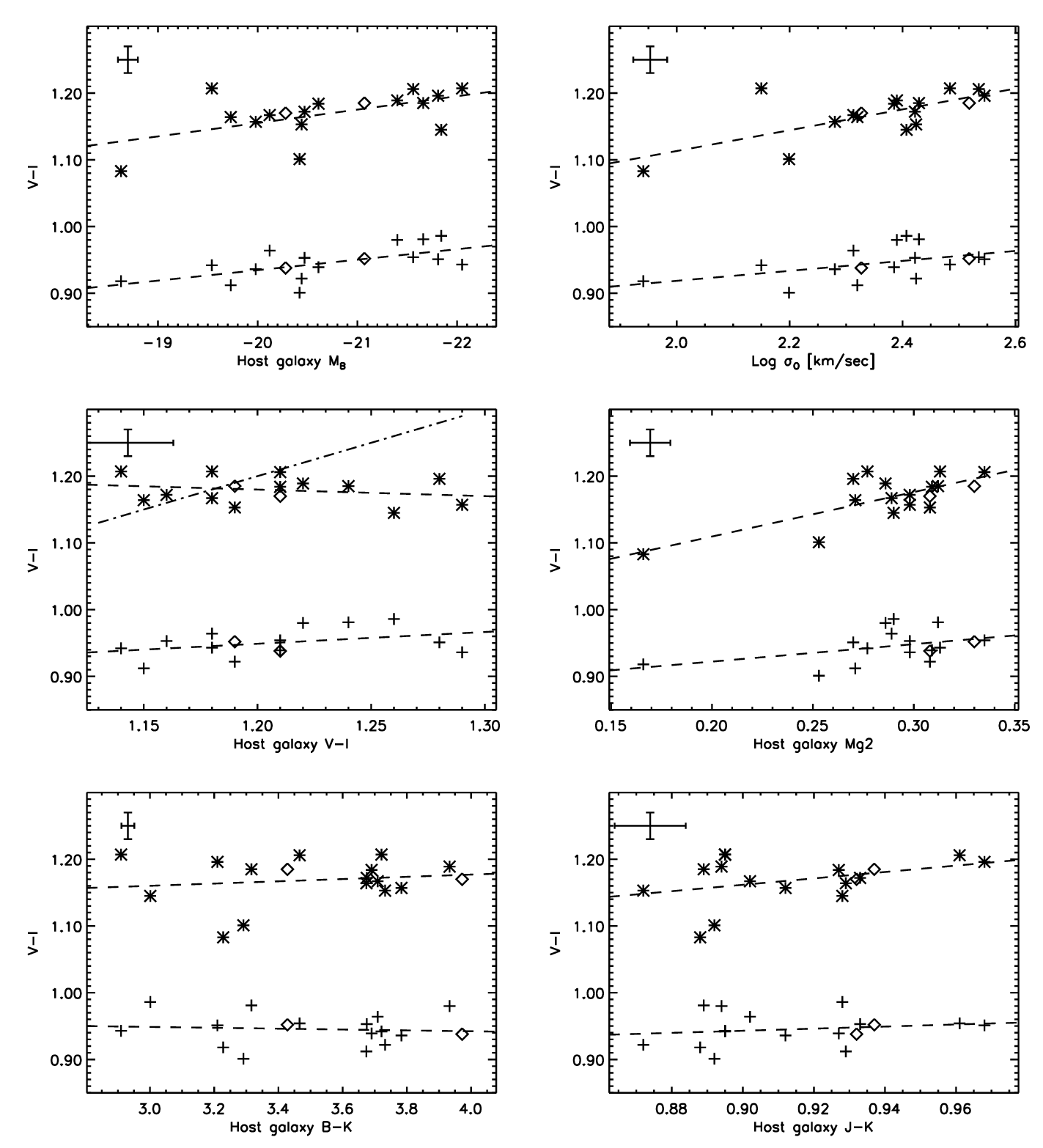}
\figcaption[Larsen.fig12.ps]{\label{fig:ccplots}
  \vio\ colors of the two peaks in the GC color distributions found by 
  the KMM test as a function of
  host galaxy $B$ magnitude, central velocity dispersion, host galaxy 
  \vio, \bko\ and \jko\ color  and Mg2 index.  Asterisks ($*$) indicate the 
  red peak, plus ($+$) markers the blue. Open diamonds indicate data for 
  NGC~1399 and NGC~1404, transformed from \bi . The dashed lines are 
  least-squares fits to the data. The dashed-dotted line in the
  host galaxy \vio\ plot corresponds to a 1:1 relation between host
  galaxy and GC colors.
}

\epsfxsize=16cm
\epsfbox{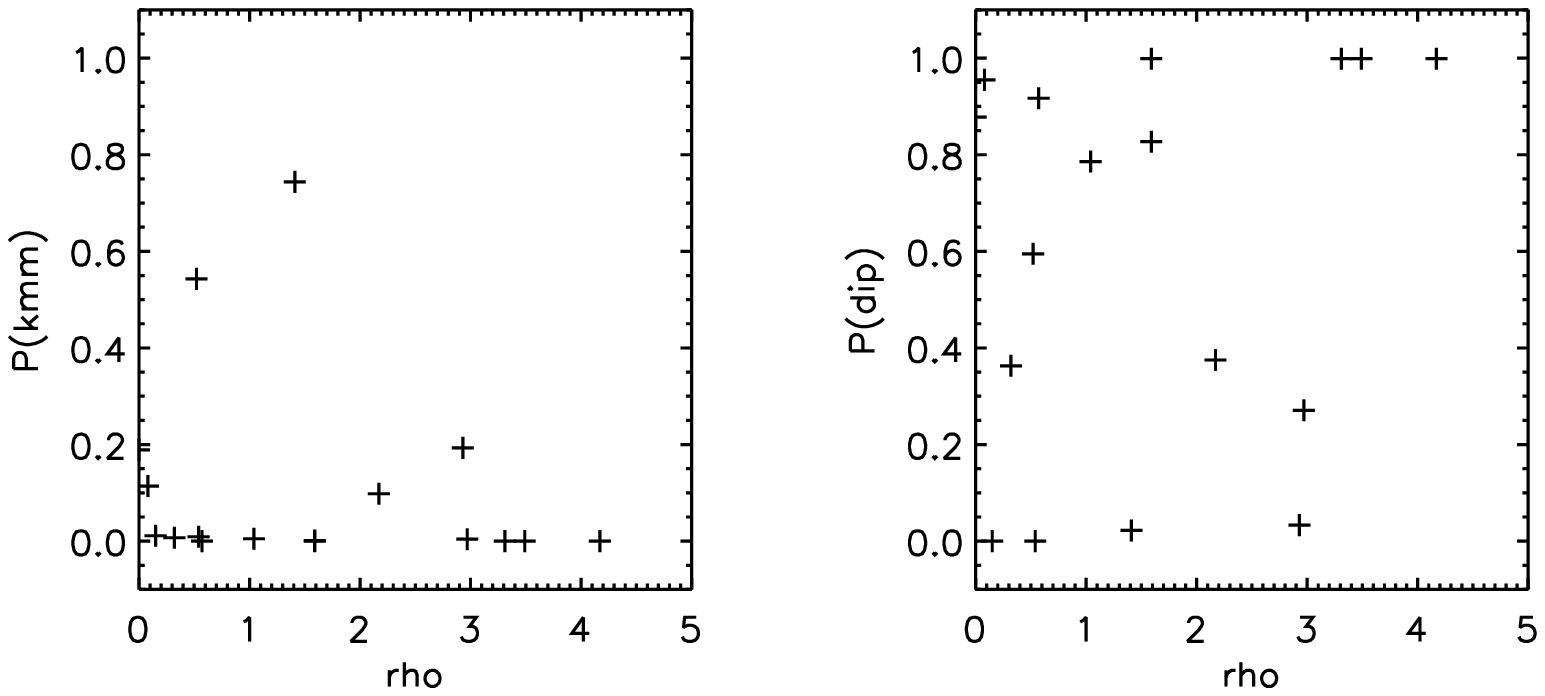}
\figcaption[Larsen.fig13.ps]{\label{fig:rho}
  Indicators of bimodality as a function of galaxy density. For $P$(kmm), a
value close to 0 indicates that two Gaussian functions are a significant
improvement relative to only one when fitting the color distribution.
For $P$(dip), a value close to 1 indicates a high probability that the
color distribution is not unimodal. 
}

\newpage

\renewcommand{\arraystretch}{.6}
\begin{table}
\begin{center}
\caption{\label{tab:gdata}Data for observations of galaxies discussed 
in this paper. Suffix '-O' indicates exposures which are offset from
the galaxy center. $^a$F450W. $^b$F547M.
} 
\begin{tabular}{llcc}\\
\tableline\tableline
Galaxy     &  PI / PID & \multicolumn{2}{c}{T$_{\rm exp}$ (sec)} \\
           &                 &  F555W & F814W \\
\tableline
NGC 524    & Brodie / 6554   &  2300  & 2300 \\
NGC 524-O  & Brodie / 6554   &  2400  & 2300 \\
NGC 1023   & Brodie / 6554   &  2400  & 2400 \\
NGC 1023-O & Brodie / 6554   &  2400  & 2400 \\
NGC 1399 & Grillmair / 5990 & 2600$^a$ & 1200 \\
NGC 1404 & Grillmair / 5990 & 2600$^a$ & 1600 \\
NGC 3115   & Faber / 5512    &  1050  & 1050 \\
NGC 3379   & Faber / 5512    &  1660  & 1200 \\
NGC 3384   & Faber / 5512    &  1050  & 1050 \\
NGC 4365   & Brodie / 5920   &  2200  & 2300 \\
NGC 4365-O & Brodie / 6554   &  2200  & 2200 \\
NGC 4406   & Faber / 5512    &  1500  & 1500 \\
NGC 4472   & Westphal / 5236 & 1800   & 1800 \\
NGC 4472-O1 & Brodie / 5920   &  2200  & 2300 \\
NGC 4472-O2 & Brodie / 5920   &  2200  & 2300 \\
NGC 4473   & Faber  / 6099   &  1800  & 2000 \\
NGC 4486   & Macchetto / 5477 & 2400  & 2400 \\
NGC 4486-O1 & Macchetto / 6844 & 2000  & 2000 \\
NGC 4486-O2 & Macchetto / 6844 & 2000  & 2000 \\
NGC 4486-O3 & Biretta / 7274 & 2200  & 2500 \\
NGC 4486-O4 & Macchetto / 6844 & 1000  & 1000 \\
NGC 4494   & Brodie / 6554   &  2400  & 1800 \\
NGC 4494-O & Brodie / 6554   &  2400  & 1600 \\
NGC 4552   & Faber  / 6099   &  2400  & 1500 \\
NGC 4594   & Faber / 5512  & 1200$^b$ & 1050 \\
NGC 4649   & Westphal / 6286 &  2100  & 2500 \\
NGC 4733   & Brodie / 6554   &  2200  & 2200 \\
\tableline
\end{tabular}
\end{center}
\end{table}

\begin{deluxetable}{rrrrrcccccc}
\rotate
\tablecaption{\label{tab:props}Host galaxy properties}
\tablecomments{Sources for the
various parameters listed in the table: Morphological types: NASA/IPAC 
Extragalactic Database, $A_B$ values: \citet{sch98}, \vi\ colors 
and $B$ magnitudes: \citet{pru98}, Mg2 indices: \citet{gol98}, 
central velocity dispersions ($\sigma_0$): the Lyon/Meudon 
Extragalactic Database.  $\rho$ = Galaxy density in Mpc$^{-3}$, from
\citet{tul88}.  Distance references:
$^1$Radial velocity, H$_0$ = 75 km / sec / Mpc.
$^2$\citet{cia91}.
$^3$\citet{els97}.
$^4$\citet{sak97}.
$^5$\citet{ne00}.
$^6$\citet{sim94}.
$^7$Assumed Virgo member.
$^8$\citet{mc93}.
$^9$\citet{for96}.
}
\tabletypesize{\footnotesize}
\tablehead{
Galaxy     & Type  & $m-M$     & $M_B$   & $A_B$ & $\log \sigma_0$ & 
   \vio\ &  \bko\ & \jko\ & Mg2 & $\rho$ \\
           &       &           &         &       &   [km/sec]      &
             &        &      &     & [Mpc$^{-3}$]
}
\startdata
NGC 524    &  S0   & 32.5$^1$  & $-21.40$ & 0.356 & $2.390 \pm 0.059$ & $1.22 \pm 0.030$ & 3.933 & 0.894 & $0.286 \pm 0.011$ & 0.15 \\
NGC 1023   &  S0   & 29.97$^2$ & $-19.73$ & 0.262 & $2.320 \pm 0.016$ & $1.15 \pm 0.014$ & 3.674 & 0.929 & $0.271 \pm 0.002$ & 0.57 \\
NGC 1399   &  cD   & 31.17$^8$ & $-21.07$ & 0.056 & $2.518 \pm 0.075$ & $1.19 \pm 0.003$ & 3.428 & 0.937 & $0.330 \pm 0.003$ & 1.59 \\
NGC 1404   &  E1   & 31.15$^8$ & $-20.28$ & 0.049 & $2.327 \pm 0.096$ & $1.21 \pm 0.003$ & 3.972 & 0.932 & $0.308 \pm 0.003$ & 1.59 \\
NGC 3115   &  S0   & 30.2$^3$  & $-20.44$ & 0.205 & $2.424 \pm 0.058$ & $1.19 \pm 0.013$ & 3.732 & 0.872 & $0.308 \pm 0.003$ & 0.08 \\
NGC 3379   &  E1   & 30.3$^4$  & $-20.12$ & 0.105 & $2.313 \pm 0.040$ & $1.18 \pm 0.001$ & 3.709 & 0.902 & $0.289 \pm 0.002$ & 0.52 \\
NGC 3384   &  S0   & 30.3$^4$  & $-19.54$ & 0.105 & $2.150 \pm 0.029$ & $1.14 \pm 0.022$ & 3.721 & 0.895 & $0.277 \pm 0.007$ & 0.54 \\
NGC 4365   &  E3   & 31.94$^5$ & $-21.66$ & 0.091 & $2.429 \pm 0.021$ & $1.24 \pm 0.002$ & 3.316 & 0.889 & $0.312 \pm 0.003$ & 2.93 \\
NGC 4406   &  E    & 31.45$^5$ & $-21.84$ & 0.128 & $2.407 \pm 0.032$ & $1.26 \pm 0.024$ & 3.001 & 0.928 & $0.290 \pm 0.003$ & 1.41 \\
NGC 4472   & E2/S0 & 30.94$^5$ & $-22.05$ & 0.096 & $2.484 \pm 0.037$ & $1.18 \pm 0.009$ & 2.910 & 0.895 & $0.313 \pm 0.002$ & 3.31 \\
NGC 4473   &  E5   & 31.07$^5$ & $-19.98$ & 0.123 & $2.280 \pm 0.023$ & $1.29 \pm 0.005$ & 3.783 & 0.912 & $0.298 \pm 0.004$ & 2.17 \\
NGC 4486   &  E0/1 & 31.15$^5$ & $-21.81$ & 0.096 & $2.545 \pm 0.029$ & $1.28 \pm 0.009$ & 3.210 & 0.968 & $0.270 \pm 0.005$ & 4.17 \\
NGC 4494   &  E1/2 & 30.88$^6$ & $-20.42$ & 0.092 & $2.199 \pm 0.029$ &  -               & 3.291 & 0.892 & $0.253 \pm 0.004$ & 1.04 \\
NGC 4552   &  E0   & 31.00$^5$ & $-20.47$ & 0.177 & $2.422 \pm 0.021$ & $1.16 \pm 0.024$ & 3.675 & 0.933 & $0.298 \pm 0.005$ & 2.97 \\
NGC 4594   &  Sa   & 29.8$^9$  & $-20.61$ & 0.221 & $2.385 \pm 0.035$ & $1.21 \pm 0.020$ & 3.690 & 0.927 & $0.309 \pm 0.010$ & 0.32 \\
NGC 4649   &  E2   & 31.06$^5$ & $-21.56$ & 0.114 & $2.535 \pm 0.028$ & $1.21 \pm 0.107$ & 3.466 & 0.961 & $0.335 \pm 0.003$ & 3.49 \\
NGC 4733   &  E    & 31.0$^7$  & $-18.63$ & 0.090 & $1.941 \pm 0.054$ &  -               & 3.228 & 0.888 & $0.166 \pm 0.005$ & - \\
\enddata
\end{deluxetable}

\begin{table}
\begin{center}
\caption{\label{tab:kmm}Results of the KMM test applied to data for
globular clusters in the galaxies. Also listed is the probability that 
the color distribution is not unimodal, calculated from the ``dip'' 
statistic ($P$(dip)).  Note $P$(kmm) close to 0 indicates a high 
probability for \emph{bi}modality, while $P$(dip) close to 0 indicates a
high probability for \emph{uni}modality.  Data for NGC~1399 and 
NGC~1404 have been converted 
from \bi\ colors using the relations in \citet{for00}.}
\begin{tabular}{lccrrcc}
\tableline\tableline
Galaxy   &  \multicolumn{2}{c}{\vio\ peaks} & \multicolumn{2}{c}{N(blue,red)} & $P$(kmm) & $P$(dip) \\ \tableline
NGC 524 & 0.980 & 1.189 & 360 & 174 & 0.011 & 0.000 \\ 
NGC 1023 & 0.912 & 1.164 & 69 & 50 & 0.000 & 0.917 \\ 
NGC 1399 & 0.952 & 1.185 & 152 & 256 & 0.000 & 0.999 \\ 
NGC 1404 & 0.938 & 1.170 & 80 & 65 & 0.001 & 0.827 \\ 
NGC 3115 & 0.922 & 1.153 & 53 & 48 & 0.114 & 0.955 \\ 
NGC 3379 & 0.964 & 1.167 & 21 & 24 & 0.543 & 0.595 \\ 
NGC 3384 & 0.942 & 1.207 & 20 & 9 & 0.009 & 0.000 \\ 
NGC 4365 & 0.981 & 1.185 & 313 & 10 & 0.193 & 0.033 \\ 
NGC 4406 & 0.986 & 1.145 & 117 & 33 & 0.744 & 0.022 \\ 
NGC 4472 & 0.943 & 1.207 & 277 & 255 & 0.000 & 0.999 \\ 
NGC 4473 & 0.936 & 1.157 & 72 & 48 & 0.098 & 0.375 \\ 
NGC 4486 & 0.951 & 1.196 & 334 & 375 & 0.000 & 0.999 \\ 
NGC 4494 & 0.901 & 1.101 & 65 & 68 & 0.005 & 0.786 \\ 
NGC 4552 & 0.953 & 1.172 & 83 & 53 & 0.004 & 0.271 \\ 
NGC 4594 & 0.939 & 1.184 & 41 & 56 & 0.007 & 0.363 \\ 
NGC 4649 & 0.954 & 1.206 & 176 & 169 & 0.000 & 0.999 \\ 
NGC 4733 & 0.918 & 1.083 & 18 & 9 & 0.189 & 0.878 \\ 
\tableline
\end{tabular}
\end{center}
\end{table}

\begin{deluxetable}{lcccrccrccc}
\rotate
\tabletypesize{\small}
\tablecaption{\label{tab:mto_var}Two-parameter $t_5$ function fits to 
the globular cluster luminosity functions.}
\tablecomments{The fits were carried out within the magnitude limits 
indicated in Fig.~\ref{fig:cmd} for the 15 galaxies with $V,I$ photometry. 
Both the turn-over magnitude (\mto ) and dispersion ($\sigma_t$) of the 
$t_5$ function were allowed to vary. For NGC~4486 (M87), only the 
central (deep) pointing has been used.  No meaningful fit could be obtained
for NGC~4733.  All magnitudes are corrected for Galactic foreground extinction.
}
\tabletypesize{\footnotesize}
\tablehead{
Galaxy  & Range &
          \multicolumn{3}{c}{Blue} & 
          \multicolumn{3}{c}{Red} & 
	  \multicolumn{2}{c}{All} \\
        & & \mto & $\sigma_t$ & N & \mto & $\sigma_t$ & N 
	& \mto & $\sigma_t$ & \dmto }
\startdata
NGC 524 & $20.0<V<26.0$ & $24.34^{+0.09}_{-0.15}$ & $1.04^{+0.13}_{-0.10}$ & 320 & $24.68^{+0.13}_{-0.15}$ & $1.04^{+0.13}_{-0.10}$ & 297 & $24.51^{+0.06}_{-0.12}$ & $1.04^{+0.08}_{-0.07}$  & $0.34^{+0.20}_{-0.18}$ \\
NGC 1023 & $19.0<V<24.0$ & $22.82^{+0.47}_{-0.48}$ & $1.80^{+0.14}_{-0.34}$ & 86 & $23.92^{+0.32}_{-0.46}$ & $1.42^{+0.27}_{-0.18}$ & 87 & $23.53^{+0.24}_{-0.38}$ & $1.62^{+0.19}_{-0.19}$  & $1.10^{+0.58}_{-0.66}$ \\
NGC 3115 & $19.0<V<24.0$ & $22.45^{+0.26}_{-0.28}$ & $1.48^{+0.29}_{-0.20}$ & 64 & $22.66^{+0.34}_{-0.32}$ & $1.49^{+0.28}_{-0.21}$ & 51 & $22.55^{+0.20}_{-0.22}$ & $1.49^{+0.24}_{-0.16}$  & $0.22^{+0.44}_{-0.41}$ \\
NGC 3379 & $20.0<V<24.0$ & $22.57^{+0.30}_{-0.29}$ & $1.01^{+0.56}_{-0.15}$ & 24 & $23.02^{+0.35}_{-0.29}$ & $1.09^{+0.43}_{-0.17}$ & 31 & $22.78^{+0.22}_{-0.21}$ & $1.05^{+0.32}_{-0.13}$  & $0.45^{+0.45}_{-0.42}$ \\
NGC 3384 & $20.0<V<24.0$ & $22.98^{+0.10}_{-0.13}$ & $0.53^{+0.21}_{-0.02}$ & 30 & $24.37^{+0.00}_{-0.59}$ & $0.89^{+0.27}_{-0.14}$ & 24 & $23.30^{+0.12}_{-0.13}$ & $0.63^{+0.12}_{-0.07}$  & $1.39^{+0.13}_{-0.60}$ \\
NGC 4365 & $20.0<V<25.0$ & $24.01^{+0.13}_{-0.15}$ & $1.12^{+0.13}_{-0.10}$ & 293 & $24.83^{+0.09}_{-0.37}$ & $1.27^{+0.15}_{-0.14}$ & 207 & $24.37^{+0.15}_{-0.16}$ & $1.22^{+0.10}_{-0.09}$  & $0.81^{+0.17}_{-0.39}$ \\
NGC 4406 & $20.0<V<25.0$ & $23.28^{+0.12}_{-0.16}$ & $1.04^{+0.17}_{-0.12}$ & 97 & $23.52^{+0.16}_{-0.18}$ & $1.09^{+0.20}_{-0.13}$ & 77 & $23.38^{+0.08}_{-0.14}$ & $1.05^{+0.11}_{-0.09}$  & $0.24^{+0.23}_{-0.22}$ \\
NGC 4472 & $20.0<V<25.0$ & $23.38^{+0.14}_{-0.17}$ & $1.09^{+0.20}_{-0.12}$ & 302 & $24.21^{+0.24}_{-0.22}$ & $1.42^{+0.17}_{-0.13}$ & 391 & $23.78^{+0.12}_{-0.15}$ & $1.29^{+0.12}_{-0.11}$  & $0.83^{+0.29}_{-0.26}$ \\
NGC 4473 & $20.0<V<25.0$ & $23.46^{+0.14}_{-0.16}$ & $0.90^{+0.19}_{-0.10}$ & 68 & $23.86^{+0.18}_{-0.22}$ & $1.09^{+0.24}_{-0.14}$ & 65 & $23.66^{+0.10}_{-0.14}$ & $1.00^{+0.12}_{-0.11}$  & $0.40^{+0.24}_{-0.26}$ \\
NGC 4486 & $20.0<V<25.0$ & $23.36^{+0.08}_{-0.12}$ & $1.25^{+0.09}_{-0.09}$ & 304 & $23.58^{+0.06}_{-0.08}$ & $1.21^{+0.05}_{-0.08}$ & 474 & $23.50^{+0.04}_{-0.08}$ & $1.22^{+0.04}_{-0.06}$  & $0.22^{+0.13}_{-0.11}$ \\
NGC 4494 & $20.0<V<24.5$ & $23.24^{+0.11}_{-0.15}$ & $0.71^{+0.21}_{-0.08}$ & 94 & $23.76^{+0.23}_{-0.20}$ & $0.72^{+0.23}_{-0.07}$ & 68 & $23.40^{+0.09}_{-0.13}$ & $0.70^{+0.13}_{-0.07}$  & $0.52^{+0.27}_{-0.22}$ \\
NGC 4552 & $20.0<V<24.5$ & $23.01^{+0.20}_{-0.22}$ & $1.34^{+0.33}_{-0.17}$ & 84 & $23.61^{+0.26}_{-0.22}$ & $1.22^{+0.24}_{-0.15}$ & 69 & $23.32^{+0.14}_{-0.19}$ & $1.33^{+0.19}_{-0.12}$  & $0.60^{+0.34}_{-0.30}$ \\
NGC 4594 & $19.0<V<23.5$ & $21.80^{+0.18}_{-0.20}$ & $0.96^{+0.37}_{-0.11}$ & 41 & $22.22^{+0.11}_{-0.13}$ & $0.80^{+0.14}_{-0.11}$ & 70 & $22.09^{+0.09}_{-0.11}$ & $0.89^{+0.12}_{-0.09}$  & $0.42^{+0.23}_{-0.22}$ \\
NGC 4649 & $20.0<V<25.0$ & $23.46^{+0.12}_{-0.14}$ & $1.33^{+0.15}_{-0.11}$ & 200 & $23.66^{+0.10}_{-0.12}$ & $1.22^{+0.11}_{-0.10}$ & 222 & $23.58^{+0.06}_{-0.10}$ & $1.28^{+0.08}_{-0.09}$  & $0.20^{+0.17}_{-0.17}$ \\
NGC~4733 & $20.0<V<25.0$ & - & - & 21 & - & - & 18 & - & - & - \\
\enddata
\end{deluxetable}

\begin{deluxetable}{lcccccccc}
\rotate
\tabletypesize{\small}
\tablecaption{\label{tab:mto_fix}One-parameter $t_5$ function fits to the 
globular cluster luminosity functions.}
\tablecomments{The fits were carried out within the magnitude limits 
indicated in Fig.~\ref{fig:cmd} for the 15 galaxies with $V,I$ photometry. 
Only the turn-over magnitude (\mto ) of the $t_5$ function was allowed to 
vary. For comparison we have also included fits to Milky Way globular
clusters brighter than $M_V=-5$, roughly corresponding to the magnitude
limit for the ellipticals.  Note: the turn-over magnitudes for the
Milky Way data are absolute.  
}
\tablehead{
Galaxy  & Range &
          \multicolumn{2}{c}{Blue} & 
          \multicolumn{2}{c}{Red} & 
	  \multicolumn{1}{c}{All} & \\
        & & \mto & N & \mto & N 
	& \mto & \dmto }
\startdata
NGC 524 & $20.0<V<26.0$ & $24.36^{+0.12}_{-0.14}$ & 320 & $24.72^{+0.13}_{-0.15}$ & 297 & $24.56^{+0.07}_{-0.13}$  & $0.36^{+0.19}_{-0.20}$ \\
NGC 1023 & $19.0<V<24.0$ & $22.45^{+0.25}_{-0.26}$ & 86 & $23.37^{+0.40}_{-0.27}$ & 87 & $22.93^{+0.20}_{-0.19}$  & $0.92^{+0.48}_{-0.37}$ \\
NGC 3115 & $19.0<V<24.0$ & $22.27^{+0.18}_{-0.20}$ & 64 & $22.43^{+0.22}_{-0.22}$ & 51 & $22.34^{+0.11}_{-0.16}$  & $0.16^{+0.30}_{-0.28}$ \\
NGC 3379 & $20.0<V<24.0$ & $22.62^{+0.35}_{-0.32}$ & 24 & $23.03^{+0.35}_{-0.28}$ & 31 & $22.82^{+0.21}_{-0.24}$  & $0.41^{+0.47}_{-0.45}$ \\
NGC 3384 & $20.0<V<24.0$ & $23.37^{+0.38}_{-0.29}$ & 30 & $24.48^{+0.00}_{-0.60}$ & 24 & $23.90^{+0.33}_{-0.27}$  & $1.11^{+0.29}_{-0.71}$ \\
NGC 4365 & $20.0<V<25.0$ & $23.98^{+0.15}_{-0.17}$ & 293 & $24.54^{+0.28}_{-0.22}$ & 207 & $24.22^{+0.13}_{-0.14}$  & $0.56^{+0.33}_{-0.27}$ \\
NGC 4406 & $20.0<V<25.0$ & $23.30^{+0.12}_{-0.18}$ & 97 & $23.52^{+0.16}_{-0.18}$ & 77 & $23.40^{+0.08}_{-0.14}$  & $0.22^{+0.24}_{-0.22}$ \\
NGC 4472 & $20.0<V<25.0$ & $23.39^{+0.13}_{-0.17}$ & 302 & $23.89^{+0.16}_{-0.17}$ & 391 & $23.65^{+0.09}_{-0.12}$  & $0.50^{+0.24}_{-0.21}$ \\
NGC 4473 & $20.0<V<25.0$ & $23.56^{+0.16}_{-0.20}$ & 68 & $23.86^{+0.20}_{-0.20}$ & 65 & $23.72^{+0.12}_{-0.16}$  & $0.30^{+0.28}_{-0.26}$ \\
NGC 4486 & $20.0<V<25.0$ & $23.30^{+0.06}_{-0.12}$ & 304 & $23.52^{+0.06}_{-0.08}$ & 474 & $23.44^{+0.04}_{-0.08}$  & $0.22^{+0.13}_{-0.10}$ \\
NGC 4494 & $20.0<V<24.5$ & $23.41^{+0.26}_{-0.23}$ & 94 & $24.15^{+0.14}_{-0.47}$ & 68 & $23.63^{+0.21}_{-0.21}$  & $0.74^{+0.27}_{-0.54}$ \\
NGC 4552 & $20.0<V<24.5$ & $22.91^{+0.15}_{-0.18}$ & 84 & $23.52^{+0.22}_{-0.20}$ & 69 & $23.19^{+0.11}_{-0.15}$  & $0.61^{+0.28}_{-0.25}$ \\
NGC 4594 & $19.0<V<23.5$ & $21.85^{+0.22}_{-0.24}$ & 41 & $22.38^{+0.19}_{-0.18}$ & 70 & $22.20^{+0.13}_{-0.15}$  & $0.53^{+0.31}_{-0.28}$ \\
NGC 4649 & $20.0<V<25.0$ & $23.34^{+0.10}_{-0.12}$ & 200 & $23.60^{+0.08}_{-0.12}$ & 222 & $23.48^{+0.06}_{-0.10}$  & $0.26^{+0.14}_{-0.16}$ \\
NGC~4733 & $20.0<V<25.0$ & - & 21 & - & 18 & - & - \\
Milky Way & $M_V<-5$ & $-7.63^{+0.15}_{-0.17}$ & 67 &
                       $-7.17^{+0.30}_{-0.35}$ & 20 &
		       $-7.55^{+0.13}_{-0.15}$ &
	       $0.46^{+0.34}_{-0.38}$ \\
\enddata
\end{deluxetable}

\begin{table}
\begin{center}
\caption{\label{tab:ldf_up} Comparison of power-law fits to the upper end
of the LFs. The fits are for $10^5 \lsun < L < 10^6 \lsun $. Also
given is a weighted mean of the power-law slopes for data with errors on
the slope less than 0.5.}
\begin{tabular}{lccc} 
\tableline\tableline
Galaxy    &  \multicolumn{3}{c}{Power-law index}                  \\ 
          &        Blue      &        Red       &        All      \\ \tableline
NGC  524  & $-1.60 \pm 0.12$ & $-1.91 \pm 0.13$ & $-1.74 \pm 0.09$ \\
NGC 1023  & $-1.70 \pm 0.26$ & $-2.49 \pm 0.37$ & $-2.02 \pm 0.21$ \\
NGC 3115  & $-1.60 \pm 0.31$ & $-1.17 \pm 0.34$ & $-1.41 \pm 0.22$ \\
NGC 3379  & $-0.80 \pm 0.68$ & $-1.23 \pm 0.60$ & $-1.00 \pm 0.45$ \\
NGC 3384  & $-2.52 \pm 0.97$ & $-1.00 \pm 1.10$ & $-2.34 \pm 0.67$ \\
NGC 4365  & $-1.69 \pm 0.11$ & $-1.79 \pm 0.16$ & $-1.73 \pm 0.09$ \\
NGC 4406  & $-1.49 \pm 0.19$ & $-1.52 \pm 0.23$ & $-1.52 \pm 0.15$ \\
NGC 4472  & $-1.66 \pm 0.15$ & $-1.87 \pm 0.14$ & $-1.78 \pm 0.10$ \\
NGC 4473  & $-1.76 \pm 0.31$ & $-2.25 \pm 0.32$ & $-2.11 \pm 0.22$ \\
NGC 4486  & $-1.65 \pm 0.11$ & $-1.79 \pm 0.10$ & $-1.74 \pm 0.08$ \\
NGC 4494  & $-2.00 \pm 0.23$ & $-2.31 \pm 0.37$ & $-2.20 \pm 0.20$ \\
NGC 4552  & $-1.43 \pm 0.21$ & $-1.56 \pm 0.29$ & $-1.61 \pm 0.17$ \\
NGC 4594  & $-2.00 \pm 0.30$ & $-1.94 \pm 0.29$ & $-1.98 \pm 0.21$ \\
NGC 4649  & $-1.59 \pm 0.15$ & $-1.68 \pm 0.16$ & $-1.64 \pm 0.11$ \\
NGC 4733  & $-1.68 \pm 0.71$ & $-1.03 \pm 0.49$ & $-1.61 \pm 0.37$ \\
Mean & $-1.66\pm0.03$ & $-1.80\pm0.06$ & $-1.74\pm0.04$ \\
Milky Way & $-2.20 \pm 0.28$ & $-1.62 \pm 0.37$ & $-2.20 \pm 0.13$ \\
\tableline
\end{tabular}
\end{center}
\end{table}

\begin{table}
\begin{center}
\caption{\label{tab:sizes}Cluster sizes for clusters brighter than
$V=24$. For the Milky Way the division between 'Blue' and 'Red' clusters 
is at [Fe/H] = $-1$. \reff = Median effective radius in pc, N = number of
clusters in each bin.}
\begin{tabular}{l*{12}{r}} 
\tableline\tableline
      &   & & \multicolumn{8}{c}{\vio\ color range} \\
              &            \multicolumn{2}{c}{Blue} & 
	                   \multicolumn{2}{c}{Red} & 
	                   \multicolumn{2}{c}{$[0.70-0.90]$} & 
	                   \multicolumn{2}{c}{$[0.90-1.05]$} &
	                   \multicolumn{2}{c}{$[1.05-1.20]$} &
	                   \multicolumn{2}{c}{$[1.20-1.45]$} \\ 
Galaxy   & \reff & N & \reff & N &
                           \reff\ & N & \reff\ & N & \reff\ & N & \reff\ & N \\
			   \tableline
NGC 524 & 2.76 & 135 & 2.59 &  92 & 2.99 &  27 & 2.76 & 108 & 2.76 &  70 & 2.07 &  22 \\ 
NGC 1023 & 1.72 &  84 & 1.65 &  71 & 1.65 &  36 & 1.79 &  48 & 1.22 &  49 & 3.73 &  22 \\ 
NGC 1399 & 1.99 & 113 & 1.37 & 212 & 2.12 &  33 & 1.99 &  80 & 1.37 & 122 & 1.37 &  90 \\ 
NGC 1404 & 1.61 &  43 & 1.61 &  47 & 1.98 &  12 & 1.36 &  31 & 1.61 &  30 & 1.48 &  17 \\ 
NGC 3115 & 2.47 &  70 & 2.23 &  56 & 2.71 &  26 & 2.39 &  47 & 2.31 &  44 & 2.23 &  16 \\ 
NGC 3379 & 3.01 &  26 & 2.25 &  31 & 3.53 &   6 & 2.93 &  20 & 2.09 &  19 & 2.76 &  12 \\ 
NGC 3384 & 2.34 &  28 & 3.34 &  15 & 2.34 &  13 & 2.34 &  15 & 3.34 &   6 & 5.34 &   9 \\ 
NGC 4365 & 2.40 & 163 & 2.88 &  79 & 2.76 &  56 & 2.31 & 107 & 2.84 &  66 & 3.44 &  13 \\ 
NGC 4406 & 4.11 &  78 & 2.98 &  57 & 4.82 &  20 & 3.83 &  58 & 3.26 &  40 & 2.84 &  17 \\ 
NGC 4472 & 1.85 & 220 & 1.46 & 249 & 1.79 &  59 & 1.91 & 161 & 1.57 & 120 & 1.23 & 129 \\ 
NGC 4473 & 2.02 &  61 & 1.90 &  44 & 2.41 &  26 & 1.79 &  35 & 1.90 &  37 & 2.38 &   8 \\ 
NGC 4486 & 2.59 & 580 & 2.04 & 757 & 3.09 & 179 & 2.47 & 401 & 2.16 & 437 & 1.85 & 320 \\ 
NGC 4494 & 2.07 &  97 & 1.82 &  61 & 2.26 &  35 & 1.96 &  62 & 1.82 &  55 & 3.38 &   6 \\ 
NGC 4552 & 1.96 &  87 & 1.97 &  80 & 2.42 &  26 & 1.96 &  61 & 1.73 &  50 & 2.07 &  30 \\ 
NGC 4594 & 1.81 &  56 & 1.43 &  84 & 2.27 &  16 & 1.81 &  40 & 1.62 &  42 & 1.17 &  42 \\ 
NGC 4649 & 1.66 & 157 & 1.30 & 175 & 2.01 &  34 & 1.66 & 123 & 1.30 &  88 & 1.42 &  87 \\ 
NGC 4733 & 2.88 &  16 & 1.40 &   7 & 3.00 &   6 & 2.88 &  10 & 1.40 &   7 & 0.0 &   0 \\ 
Milky Way & 3.24 & 96 & 2.29  & 39 & 3.28 & 60 & 3.03 & 36 & 2.37 & 24 & 2.07 & 15 \\
\tableline
\end{tabular}
\end{center}
\end{table}

\begin{table}
\begin{center}
\caption{\label{tab:sz_r}Cluster sizes and \vi\ colors for blue and red 
clusters (brighter than $V=24$) for different radial bins in selected 
cluster-rich galaxies.}
\begin{tabular}{lccrrcc} 
\tableline\tableline
Galaxy   & \multicolumn{2}{c}{\reff (med,pc)} & N(blue) & N(red) & \multicolumn{2}{c}{\vio }\\
$r$ in arcmin          & Blue  & Red    &       &    & Blue & Red \\ \tableline
\tableline
NGC 4472 \\
$0.0<r<0.5$ &  2.27 & 1.66 & 18 & 25 & 0.96 & 1.20 \\ 
$0.5<r<1.0$ &  1.91 & 1.23 & 45 & 46 & 0.97 & 1.19 \\ 
$1.0<r<1.5$ &  1.91 & 1.40 & 40 & 63 & 0.94 & 1.20 \\ 
$1.5<r<2.0$ &  1.68 & 1.57 & 38 & 43 & 0.95 & 1.24 \\ 
$2.0<r<3.0$ &  1.85 & 1.23 & 43 & 46 & 0.93 & 1.22 \\ 
$3.0<r<4.0$ &  1.79 & 1.96 & 36 & 26 & 0.94 & 1.19 \\ 
NGC 4486 \\
$0.0<r<0.5$ &  2.84 & 2.00 & 70 & 168 & 0.95 & 1.19 \\ 
$0.5<r<1.0$ &  2.35 & 1.85 & 150 & 242 & 0.96 & 1.19 \\ 
$1.0<r<1.5$ &  2.66 & 2.16 & 169 & 190 & 0.93 & 1.18 \\ 
$1.5<r<2.0$ &  2.84 & 1.98 & 83 & 81 & 0.94 & 1.19 \\ 
$2.0<r<3.0$ &  2.66 & 2.53 & 81 & 64 & 0.93 & 1.14 \\ 
$3.0<r<4.0$ &  2.41 & 2.10 & 47 & 22 & 0.93 & 1.22 \\ 
$4.0<r<6.0$ &  2.35 & 2.59 & 40 & 22 & 0.94 & 1.17 \\ 
$6.0<r<9.0$ &  2.59 & 2.84 & 35 & 14 & 0.92 & 1.20 \\ 
NGC 4365 \\
$0.0<r<0.5$ &  3.56 & 2.88 & 34 & 13 & 0.95 & 1.16 \\ 
$0.5<r<1.0$ &  1.78 & 2.84 & 44 & 18 & 0.92 & 1.04 \\ 
$1.0<r<1.5$ &  1.78 & 1.95 & 44 & 20 & 0.91 & 1.10 \\ 
$1.5<r<2.5$ &  3.91 & 3.20 & 21 & 13 & 1.01 & 1.32 \\ 
$2.5<r<3.5$ &  5.15 & 5.51 & 17 & 13 & 0.92 & 1.14 \\ 
\tableline
\end{tabular}
\end{center}
\end{table}

\begin{table}
\begin{center}
\caption{\label{tab:sz_r_mw}Sizes for metal-poor and metal-rich GCs in
the Milky Way in three radial bins.}
\begin{tabular}{lccrr}
\tableline\tableline
    &  \multicolumn{2}{c}{\reff (med,pc)}  & N($\feh<-1$) & N($\feh>-1$) \\
    &  $\feh<-1$ &  $\feh>-1$ & & \\
\tableline
 $R_g < 2$ kpc      & 2.33 & 1.98 & 13 & 14 \\
 $2 < R_g < 5$ kpc  & 2.56 & 2.38 & 19 & 15 \\
 $5 < R_g < 10$ kpc & 3.03 & 2.52 & 20 &  7 \\
\tableline
\end{tabular}
\end{center}
\end{table}

\begin{table}
\begin{center}
\caption{\label{tab:ccorr}Slopes, Spearman correlation coefficients ($\rho$) 
and the probability that correlations are present ($P_C$) for various
 host galaxy parameter vs.\ GC color relations.}
\begin{tabular}{lrrrrrr}
\tableline\tableline
	   & \multicolumn{3}{c}{Blue GCs} & \multicolumn{3}{c}{Red GCs} \\
   & \multicolumn{1}{c}{Slope}  &   \multicolumn{1}{c}{$\rho$}   & $P_C$ &
     \multicolumn{1}{c}{Slope}  &   \multicolumn{1}{c}{$\rho$}   & $P_C$ \\
\tableline
$M_B$ & $-0.016\pm0.005$ & $-0.64$ & 98.9\% & $-0.020\pm0.008$ & $-0.45$ & 92.6\% \\
$\log\sigma_0$ & $0.075\pm0.035$ & $0.51$ & 95.8\% & $0.156\pm0.042$ & $0.53$ & 96.5\% \\
\vio\ & $0.176\pm0.120$ & $0.39$ & 85.6\% & $-0.098\pm0.120$ & $0.03$ &  9.6\% \\
Mg2 & $0.262\pm0.147$ & $0.38$ & 86.8\% & $0.667\pm0.158$ & $0.43$ & 91.6\% \\
\bko\ & $-0.007\pm0.020$ & $-0.12$ & 37.6\% & $0.017\pm0.028$ & $-0.00$ &  1.2\% \\
\jko\ & $0.158\pm0.226$ & $0.24$ & 66.1\% & $0.480\pm0.306$ & $0.34$ & 82.3\% \\
\tableline
\end{tabular}
\end{center}
\end{table}

\end{document}